\UseRawInputEncoding

\documentclass[
aps,
prb,
twocolumn,
superscriptaddress,%
showpacs,%
preprintnumbers,%
byrevtex,%
floatfix%
]{revtex4}

\usepackage{dcolumn}
\usepackage{bm}
\usepackage{ifthen}
\usepackage[usenames]{color}
\usepackage{amssymb}
\usepackage{amsfonts}
\usepackage{amsmath}
\usepackage[dvips]{graphicx}
\usepackage{graphpap}

\begin{document}

\newcommand{\alat}[0]{a_{\rm lat}}
\newcommand{\clat}[0]{c_{\rm lat}}

\date{\today}
\title{First-principles calculation of electroacoustic properties of wurtzite (Al,Sc)N }

\author{Daniel F. Urban}
\email{daniel.urban@iwm.fraunhofer.de} 
\affiliation{Fraunhofer Institute for Mechanics of Materials IWM, W\"ohlerstr.\ 11, 79108 Freiburg, Germany}

\author{Oliver Ambacher}
\affiliation{Fraunhofer Institute for Applied Solid State Physics IAF, Tullastrasse 72, 79108 Freiburg, Germany}
\affiliation{INATECH--Department of Sustainable Systems Engineering, Albert-Ludwigs-Universit\"at Freiburg, Emmy-Noether-Str. 2, 79110 Freiburg, Germany}

\author{Christian Els\"asser}
\affiliation{Fraunhofer Institute for Mechanics of Materials IWM, W\"ohlerstr.\ 11, 79108 Freiburg, Germany}
\affiliation{Freiburg Materials Research Center (FMF), Albert-Ludwigs-Universit\"at Freiburg, Stefan-Meier-Str. 21, 79104 Freiburg, Germany}

\begin{abstract}
We study the electroacoustic properties of aluminum scandium nitride crystals  Al$_{1-x}$Sc$_x$N with the metastable wurtzite structure by means of first-principles calculations based on density functional theory. We extract the material property data relevant for electroacoustic device design, namely the full tensors of elastic and piezoelectric constants. Atomistic models were constructed and analyzed for a variety of Sc concentrations $0\le x\le50$\%. The functional dependence of the material properties on the scandium concentration was extracted by fitting the data obtained from an averaging procedure for different disordered atomic configurations. We give an explanation of the observed elastic softening and the extraordinary increase in piezoelectric response as a function of Sc content in terms of an element specific analysis of bond lengths and bond angles. 
\end{abstract}

\maketitle
\section{Introduction}

Metastable aluminum scandium nitride (denoted by Al$_{1-x}$Sc$_x$N or (Al,Sc)N in the following) with the wurtzite-type crystal structure belongs to the class of polar-piezoelectric materials. It can be synthesized up to approximately $x\simeq0.41$ by reactive DC or RF magnetron sputtering and is known to have outstanding electroacoustic properties, surpassing reported values for all other group-III
nitrides.\cite{Matloub2011,Umeda2013,Konno2014,Parsapour2018,Kurz2019,Ichihashi2014,Ichihashi2016,
Carlotti2017,Lu2018,Mertin2018} 
The electroacoustic properties of (Al,Sc)N depend strongly on the Sc concentration, which offers an additional degree of freedom for adjusting, e.g., the phase velocity and electromechanical coupling in the design of resonator devices.\cite{Wingqvist2010,Feil2019}

Experimentally, elastic and piezoelectric tensor components have been acquired from acoustic resonance experiments\cite{Matloub2011,Umeda2013,Konno2014,Parsapour2018,Kurz2019} or Brillouin scattering\cite{Ichihashi2014,Ichihashi2016,Carlotti2017} on (Al,Sc)N thin films with usually low Sc concentrations. Recently, the full set of electroacoustic properties was determined experimentally from Al$_{1-x}$Sc$_x$N thin films in a large range of compositions, $0\le x\le 0.32$, from the same material source using Rayleigh-type waves in surface acoustic wave (SAW) resonators.\cite{Kurz2019}
The elastic and piezoelectric properties of (Al,Sc)N have also been obtained theoretically by means of  density functional theory (DFT).\cite{Tasnadi2010, Hoglund2010, Zhang2013, Caro2015} 
However, the quantitative computation of material properties of randomly disordered alloys remains a difficult and time consuming task and raises principal conceptual questions.

We present here a comprehensive study of the electro\-acoustic properties of aluminum scandium nitride. The full set of material property data relevant for electroacoustic device design, namely the full tensors of elastic and piezoelectric constants, are computed by means of atomistic simulations based on DFT. A combinatorial approach is chosen which includes a large number of structure models and allows for a statistical analysis of the microscopic structural parameters, namely bond lengths and bond angles. This analysis gives insight into the microscopic origin of the observed highly non-linear dependence of the most relevant elastic and piezoelectric constants as a function of Sc content.

The paper is divided into three major sections that present and discuss the results for the structural parameters (Sec.\ \ref{sec:model_structures}), the elastic tensor (Sec.\ \ref{sec:Elastic_tensor}), and the piezoelectric tensor (Sec.\ \ref{sec:Piezo_tensor}) of Al$_{1-x}$Sc$_x$N. A summary is given in Sec.\ \ref{sec:summary}.

\section{Methodology}

\subsection{Modelling of disorder}
Mixed crystals and random alloys are characterized by two (or more) atomic species sharing the same (sub)lattice without giving rise to long range order. The modeling using DFT simulations of bulk materials requires the use of supercells containing a limited number of atoms (typically from a few tens to a few hundreds) in combination with periodic boundary conditions. Therefore, every structure model of finite size will be biased by the choice of the specific disorder representation. One possibility to address this problem is to construct specific model structures with as many atoms as possible (limited by computational resources) for which site occupation correlations are minimized. Such representative structure models are referred to as special quasirandom structures (SQS).\cite{Zunger1990} 

The advantage of the SQS approach is that it reduces the propensity of artifacts generated by a special (e.g. highly symmetric) local atomic environment. The drawback is the need of rather big supercells. Therefore, often only a single SQS supercell is examined as being representative for a specific atomic composition ratio. However, a single SQS structure is not a unique representation of a disordered alloy, even for large supercells. Moreover, the assumption of a purely random distribution of the different elements on the lattice is not a unique choice and in general there is the possibility of short range correlations.

A second possibility to model mixed crystals is to take a combinatorial approach and to study the large variety of different realizations belonging to the same chemical composition and to extract physical properties by an appropriate averaging procedure. Here, the variation in the physical property of interest may be traced back to specific local atomic environments. This opens up the possibility of gaining a deeper understanding of the interplay of structural elements and material properties. Both theoretical approaches, the SQS-method and the combinatorial approach, have advantages and disadvantages and complement each other. In this work we take the second approach.

In order to model the (Al,Sc)N alloy with random distribution of Al and Sc atoms on the metal sublattice we used various supercell representations with varying Sc content. It is necessary to find a good compromise for the choice of the supercell models regarding the following three aspects: (i) A reasonable number of representative structure models for each Sc content are required in order to account for the influence of different local atomic configurations and to obtain reasonable statistics with respect to the local atomic environments considered. (ii) The supercell size should be large enough to avoid serious finite size effects. (iii) The supercell size should be small enough to keep the computational resources for a precise evaluation of the material parameters at a tractable  level. We have therefore chosen supercell models which contain 36 atoms in order to adequately address the above mentioned three points (see Sec. \ref{sec:model_structures}).

\subsection{Computational settings}
The calculation of elastic and piezoelectric constants was carried out using the PWscf code of the Quantum Espresso (QE)  software package \cite{PWSCF, Giannozzi2009}  using the GGA-PBE functional for exchange-correlation. The wave functions of the valence electrons are represented by a plane-waves basis set with a cutoff energy of 55 Ry (1 Rydberg $\approx 13.606$ eV), and the electron density and effective Kohn-Sham potential by discrete Fourier series with a cutoff energy of 440 Ry. The interactions of valence electrons with the atomic nuclei and core electrons are described by pseudopotentials taken from the open-source 
Standard Solid State Pseudopotentials (SSSP) library.\cite{Prandini2018, Lejaeghere2016} Here, ultrasoft pseudopotentials were chosen for N and Sc, while the pseudopotential for Al is of PAW type. Brillouin-zone integrals for the 36-atom supercells were evaluated on a Monkhorst-Pack mesh of 3x3x6 k-points with a Gaussian smearing of 0.01 Ry. The convergence threshold was set to 10$^{-5}$ Ry for the total energy and to 10$^{-4}$ Ry/Bo (1 Bohr = 0.529\AA) for the forces on atoms. Elastic stresses and interatomic forces were relaxed using the Broyden-Fletcher-Goldfarb-Shanno (BFGS) algorithm.

\section{Structure of (Al,Sc)N}
\label{sec:model_structures}

This section introduces our representation of disordered model structures for (Al,Sc)N. The energy criterion guiding the choice of model structures is presented in Sec. \ref{subsec:low_E_SCmodels} and the evaluation of lattice parameters in Sec.\ \ref{subsec:lattice_param}.  An explanation of the origin of the observed highly anisotropic change of lattice parameters in terms of bond lengths and bond angles is given in Sec.\ \ref{subsec:origin_anisotropy}. 

\begin{figure}[]
\begin{center}
\setlength{\unitlength}{1mm}
\begin{picture}(85,38)(0,0)
\put(0,0){\includegraphics[width=3cm,draft=false]{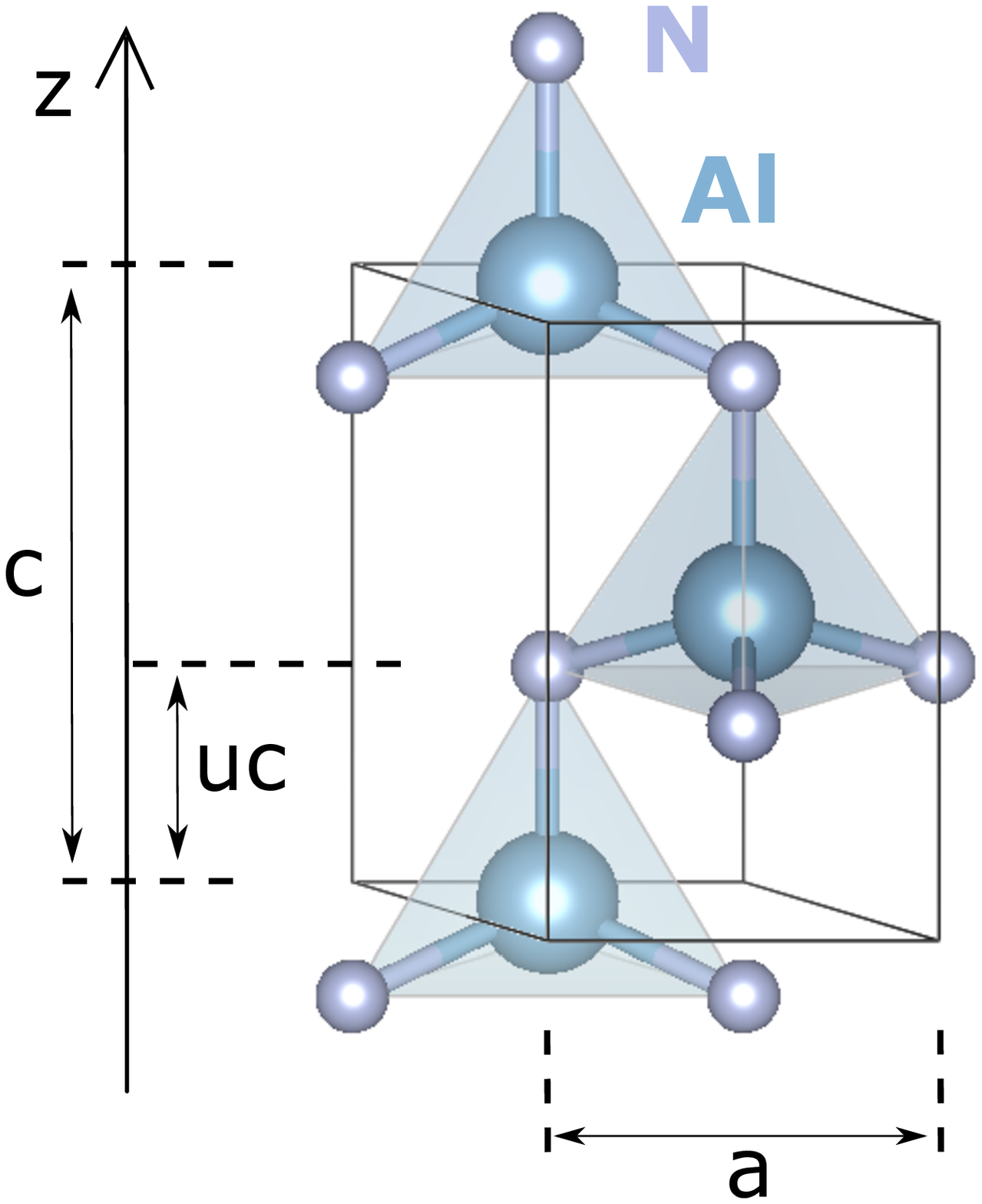}}
\put(40,0){\includegraphics[width=4.5cm,draft=false]{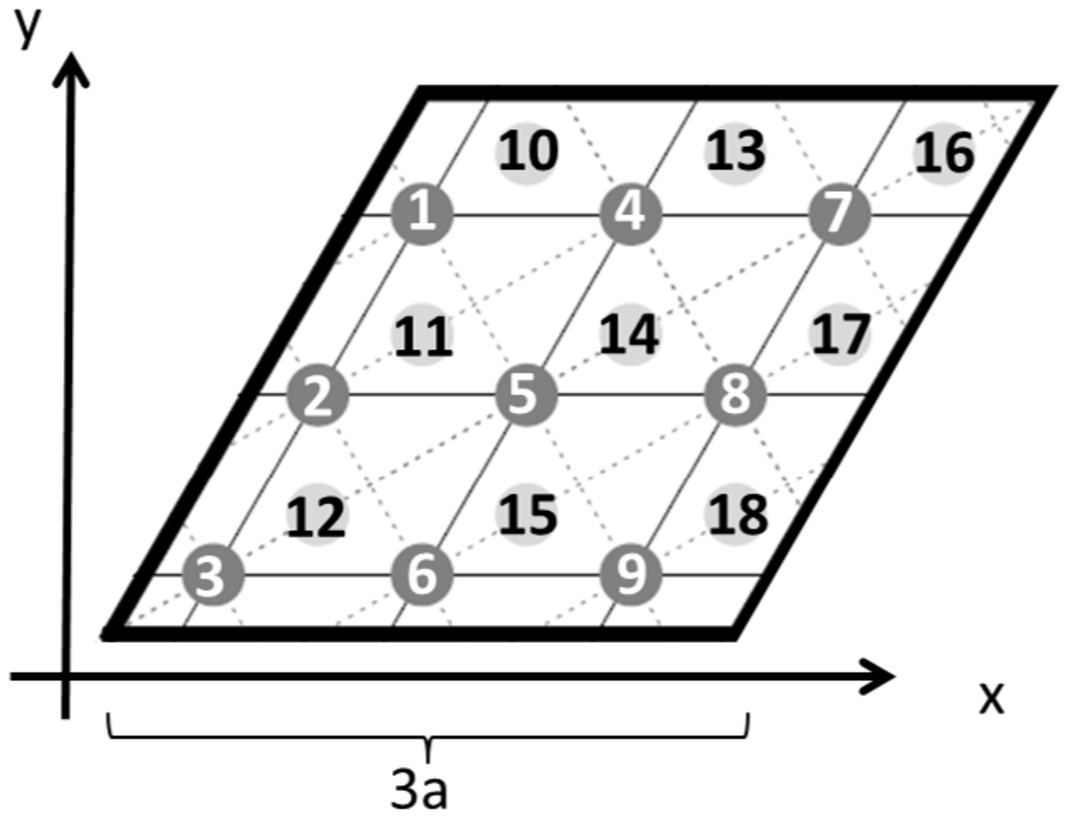}}
\end{picture}
\caption{
Left: The AlN unit cell of the wurtzite crystal structure which is characterized by the hexagonal lattice parameters $\alat$ and $\clat$ and one internal structure parameter $u$. The latter determines the relative shift of the N sublattice with respect to the Al sublattice. Right: Top view on the 36-atom supercell illustrating the two shifted hexagonal lattices of the wurtzite structure and the positions of the 18 metal atoms.
\label{fig:wurtzite}}
\end{center}
\end{figure}     

\subsection{Low-energy supercell realizations of (Al,Sc)N}
\label{subsec:low_E_SCmodels}
In this work we consider supercell models which contain 36 atoms and are built from 3x3x1 AlN wurtzite unit cells with individual Al atoms being substituted by Sc atoms, c.f.\ Fig.\ \ref{fig:wurtzite}. 
The choice of the representative set of disorder configurations is guided by comparing the DFT total energies of the various possible atomic configurations at fixed Al:Sc ratio for this supercell size. These total energies are the ground-state energies of the structurally optimized supercell models which are obtained by relaxation of the atom positions and the cell shape (i.e. lattice parameter $\alat$ and $\clat$) to zero elastic stress and zero atomic forces. In other words, the lattice constants and atomic coordinates are determined such that the total energy is minimal for the given distribution of Al and Sc atoms on the metal sublattice in the supercell.

\begin{figure}[]
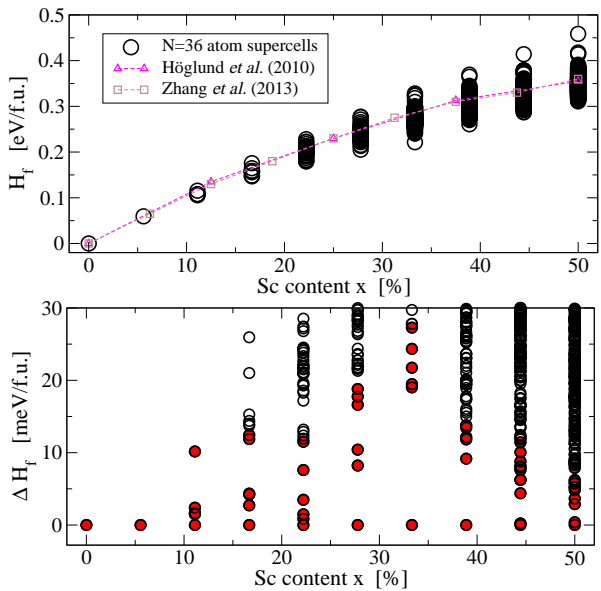

\begin{center}
\includegraphics[width=0.91\columnwidth]{Fig2a_Emix_vs_x.eps}
\includegraphics[width=0.9\columnwidth]{Fig2b_relEform_vs_x.eps}
\caption{Formation enthalpies of metastable wurtzite (Al,Sc)N.
Upper panel: Formation enthalpies $H_f$ with respect to the binary nitride phases, evaluated for all the inequivalent 36-atom supercells that are combinatorially possible. Data taken from Refs.  [\onlinecite{Zhang2013, Hoglund2010}] are shown for comparison. Lower panel: Relative formation energy $\Delta H_f$ zoomed in on the low-energy range. Here, $\Delta H_f$ is defined as the difference in formation energy with respect to the lowest-energy structure at each given $x$. Red circles mark the structures for the detailed analysis which subsequently enter the fitting of the functional dependence of the material properties on the Sc content. 
\label{fig:Eform}}
\end{center}
\end{figure}  

Note that we have kept the hexagonal symmetry for our supercell models in the structural optimization, i.e. we optimized the cell volume and c/a ratio while keeping the angles in the hexagonal system fixed.
Microscopically there will be local shear strains, because the hexagonal supercell symmetry is broken in most cases of disordered arrangements of Sc atoms on Al sites.
However, due to the periodic boundary conditions used in the DFT simulations, we always have a structure with identical atomic arrangements in the neighboring supercells. This is not the case in the macroscopic experimental realization of a (truly) disordered material. Here, strain effect average out and  the wurtzite crystal structure is observed experimentally with zero off-diagonal elements in the lattice matrix. Therefore, by keeping the hexagonal shape fixed for our supercell models, we eliminate all microscopic broken-symmetry effects in the simulations like in the experiments.

We have screened the ground-state energy (total energy) for all the Al$_{18-n}$Sc$_n$N$_{18}$ supercells, which are combinatorically possible. Symmetry inequivalent structures were generated using the software package SOD.\cite{Grau2007} While this results in only a small number of structures for small numbers $n$ of Sc atoms in the 36 atom supercell, namely 1, 5 and 14 inequivalent structures for $n=1$, 2, and 3, respectively, this number grows considerably at higher Sc content. For $n=4$, 5, 6, 7 and 8 there are 46, 99, 219, 336 and 475 structures and for a Sc content of 50\% the wurtzite structure has 504 inequivalent possibilities for how to distribute the 9 Sc atoms on the 18 metal sublattice sites.

The upper panel of Fig.\ \ref{fig:Eform} displays the computed formation enthalpy $H_f$ for all 36-atom supercell realizations as a function of the Sc content $x=n/18$. Here, $H_f$ is defined
with respect to the two binary nitride phases, namely wurtzite AlN and cubic rocksalt-type ScN, and is computed as weighted difference in total energies.
Wurtzite (Al,Sc)N is known to be thermodynamically metastable (i.e. $H_f >0$) and can only be stabilized experimentally as thin films. Other numerical data taken from literature \cite{Zhang2013, Hoglund2010} are shown for comparison. Note that the respective authors have used 128-atoms SQS supercells with one specifically selected distribution of Sc atoms for each considered value of Sc content.

\begin{figure}[]
\begin{center}
\includegraphics[width=\columnwidth]{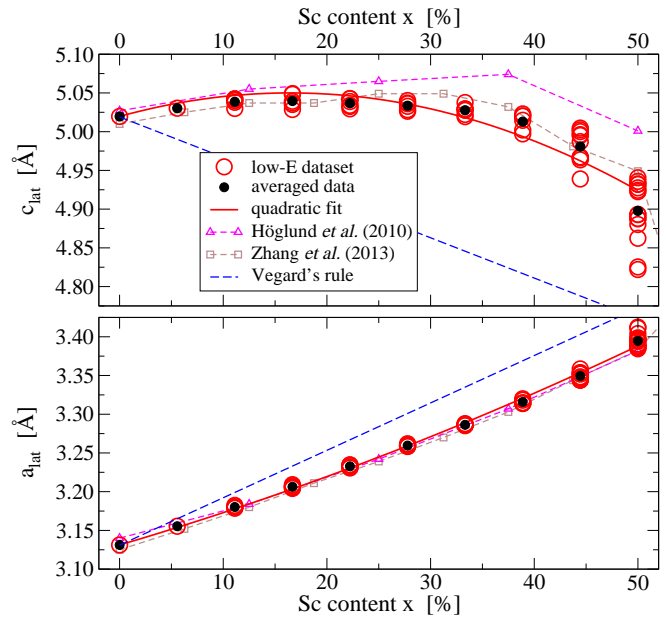}
\caption{
Lattice parameters $\alat$ and $\clat$ as a function of the Sc content $x$. The results of the quadratic fitting to the dataset are shown as solid lines. Results for $N=128$ SQS supercells taken from literature\cite{Zhang2013, Hoglund2010} are shown for comparison. The dashed blue line indicates Vegard's rule, i.e.\ the linear interpolation between the properties of the two binary compounds AlN and hexagonal ScN. 
\label{fig:alat_clat}}
\end{center}
\end{figure} 

The lower panel of Fig.\ \ref{fig:Eform} shows a close-up on the low-energy range and plots the relative formation energy $\Delta H_f$, which is the energy difference in formation energy with respect to the lowest energy structure at each given $x$. 
For (i) the further analysis of the structural parameters, (ii) the evaluation of elastic and piezoelectric tensors, and (iii) the extraction of their functional dependence on $x$, we have selected a set of lowest-energy structures for each Al:Sc ratio considered. These structures are marked by red symbols in Fig.\ \ref{fig:Eform}. Naturally, for $n=0$ and $1$ there is only one model structure each and for $n=2$ we have taken all five available structures. For $n=3$, 4, 5, 6, and 7 we have chosen the six lowest-energy structures, each. As the energetical separation between the individual structures becomes very small at large Sc content we have selected 11 and 15 sample structures for $n=8$ and 9, respectively.

Note that in the thin-film-deposition synthesis of such disordered semiconductor alloys it is equally unlikely that the resulting film yields a super-structure corresponding to the lowest-energy structure model or that it has a completely randomly disordered structure. Therefore, we decided to analyze not only the particular lowest energy structure at each given Al/Sc ratio but a larger ensemble of structures in the given energy range (shown in the lower panel of Fig.\ \ref{fig:Eform}) with equal statistical weights.


\subsection{Results: Structural parameters of (Al,Sc)N}
\label{subsec:lattice_param}

Our results for the optimized lattice parameters $\alat$ and $\clat$ are shown in Fig.\ \ref{fig:alat_clat} (see Fig.\ \ref{fig:wurtzite} for the definition of these quantities). They are found to be in very good agreement with the data from Refs.\ [\onlinecite{Hoglund2010, Zhang2013}] and indicate that our combinatorial supercell approach yields equivalent results to theirs obtained with the much larger 128-atoms cells. Naturally, the $N=36$ supercells show a dependency on the specific distribution of Sc atoms in the cell, which becomes more pronounced at larger $x$. The authors of Refs.\ [\onlinecite{Hoglund2010, Zhang2013}] each have used one specifically selected 128-atoms supercell obtained via the SQS-approach for each Sc-concentration $x$, at the expense of using a relatively low plane-wave energy cut-off and k-points density. It is interesting to note that, although the SQS are designed to distribute the Sc atoms in an uncorrelated manner as well as possible, the results of Refs.\ [\onlinecite{Hoglund2010, Zhang2013}] obtained with two distinct disorder representations differ noticeably and in a similar range as our data. 

In order to extract the functional dependence of the calculated structural parameters on the Sc content we have fitted a quadratic function to our selected set of data points. Therefore we have first averaged the datapoints separately for each $x$ value (c.f.\ black filled circles in Fig.\ \ref{fig:alat_clat}) and then applied a least-squares fitting procedure where the value of the quadratic fitting function at $x=0$ was kept fixed to the respective data point. The results for the variation in lattice parameters with Sc content $x$ are  
\begin{eqnarray}
\alat(x) &=& 3.131 \left(1 + 0.126\,x + 0.077\,x^2 \right)\,{\rm \AA},\\
\clat(x) &=& 5.020 \left(1 + 0.073\,x - 0.223\,x^2 \right)\,{\rm \AA}.
\end{eqnarray}
The corresponding quadratic fit of the respective cell volume data yields the mass density
\begin{eqnarray}
\rho(x) &=& 3.194 \left(1 + 0.108\,x + 0.030\,x^2 \right)\,{\rm g/cm^3}.
\end{eqnarray}

\begin{table}[]
\begin{center}
\begin{tabular}{l c c c c c c c}
& $\alat$ [\AA] & $\clat$ [\AA] & $u$ & $\alpha$ [$^\circ$] & $\beta$ [$^\circ$] & $\ell_{ab}$ [\AA] & $\ell_{c}$ [\AA]\\
\hline
w-AlN 	     & 3.131 & 5.019 & 0.381 & 108.2 & 110.7 & 1.903 & 1.915 \\
h-ScN 	     & 3.723 & 4.498 & 0.5   & 90    & 120   & 2.149 & 2.249 \\
average      & 3.437 & 4.759 & 0.441 & 99.2  & 115.4 & 2.026 & 2.082
\end{tabular}
\caption{
Equilibrium lattice parameters $\alat$ and $\clat$, internal parameter $u$, bond angles $\alpha$ and $\beta$,
and bond lengths $\ell_{ab}$ and $\ell_{c}$ for wurtzite AlN and hexagonal ScN from
DFT calculations. (See Figs.\ \ref{fig:wurtzite} and \ref{fig:geom_def} for the geometric definitions.) The third line gives the arithmetic mean of the values of \mbox{w-AlN} and \mbox{h-ScN}.
Note that the equilibrium lattice parameter of cubic ScN obtained
with comparable numerical settings is 4.509~\AA, yielding a ScN bond length of 2.254~\AA. 
\label{tab:binary_phases}}
\end{center}
\end{table}

The evolution of the lattice parameters with increasing Sc content is found to be highly anisotropic in agreement with experimental results.\cite{Kurz2019} The lattice parameter $\alat$ grows essentially linearly following Vegard's rule
$\alat(x)\sim (1-x) a_{\rm{AlN}} + x\, a_{\rm{ScN}}$,  
where $a_{\rm{AlN}}$ and $a_{\rm{ScN}}$ are the equilibrium lattice parameters of wurtzite AlN and hexagonal ScN. Note that the latter structure corresponds to wurtzite with the internal parameter set to $u=0.5$ and is a hypothetical crystal structure for ScN.\cite{Farrer2002,Tasnadi2010} The equilibrium crystal structure of ScN is cubic rocksalt. However, the Sc--N bond length in hexagonal ScN is very close to the one of cubic rocksalt ScN. 

In contrast, the lattice parameter $\clat$ changes on a much smaller scale and remains almost constant for a wide range of Sc content. This behavior is very different from the other mixed wurtzite nitrides like (Al,Ga)N or (Al,In)N (see e.g.\ Ref.\ \onlinecite{Dridi2003} and Refs.\ therein) and has so far not yet been fully explained. Our set of model structures, however, allows us to clarify the origin  of the anisotropic dependence on the Sc content, as discussed in the following section.

\begin{figure}[b]
\begin{center}
\includegraphics[width=0.48\columnwidth,draft=false]{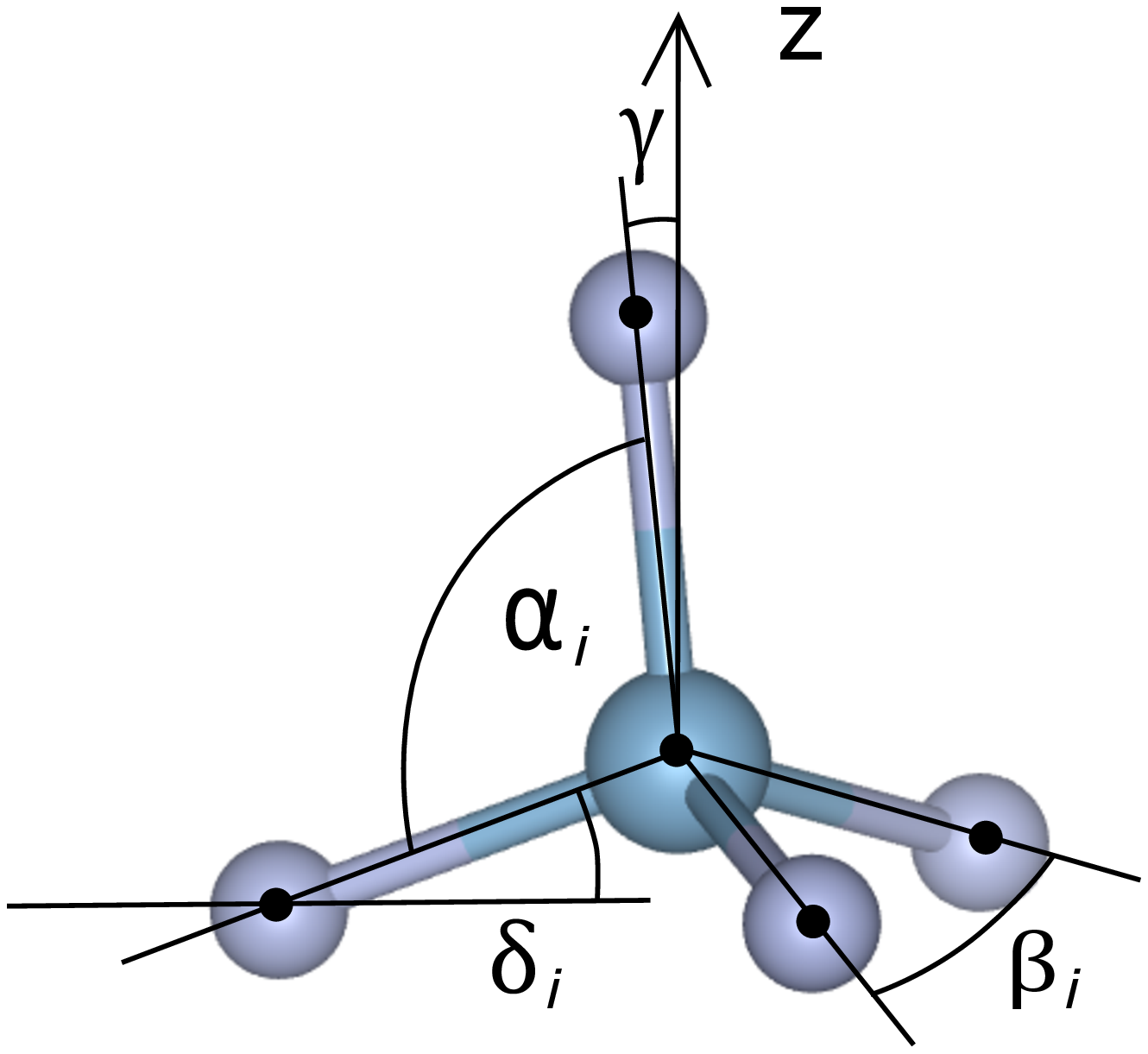}
\includegraphics[width=0.48\columnwidth,draft=false]{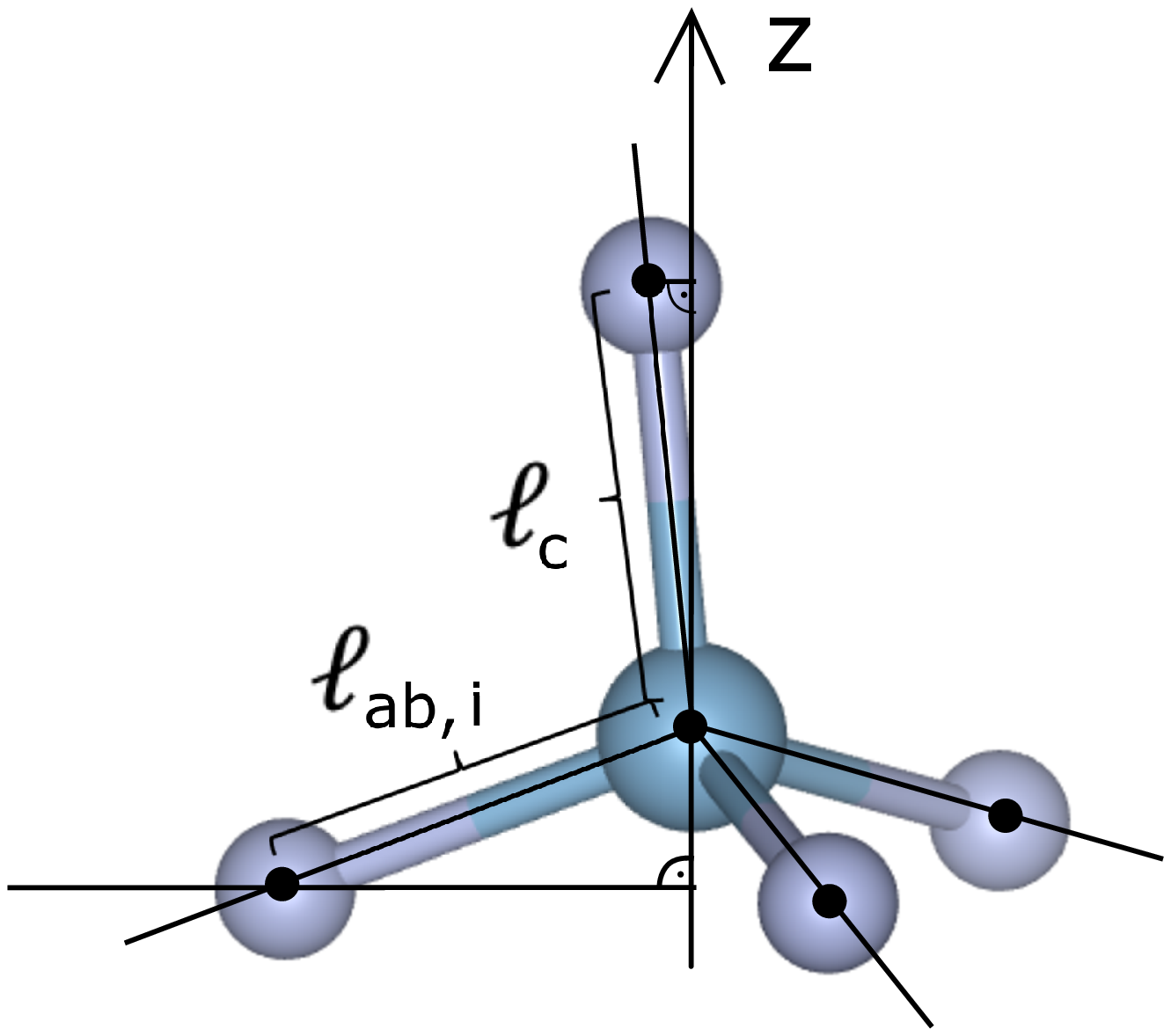}
\caption{
Definition of the bond angles (left) and bond lengths (right) for the tetrahedral MN$_4$ structural unit. Here M refers to a metal atom, M=Al or Sc.   
While the three angles $\alpha_i$ ($i=1,2,3$) for all three nitrogen atoms in the basal plane are equal in perfect wurtzite AlN, they individually differ in the mixed, symmetry-broken case of  (Al,Sc)N compounds. The same applies for the angles $\beta_i$ and $\delta_i$ and the bond lengths $\ell_{ab,i}$. 
\label{fig:geom_def}}
\end{center}
\end{figure}

\subsection{Microscopic origin of anisotropic change of lattice parameters}
\label{subsec:origin_anisotropy}

The almost perfect tetrahedra AlN$_4$ of nearest-neighbor atoms in bulk wurtzite AlN are characterized by two bond lengths $\ell_c$ and $\ell_{ab}$ for the Al--N bond parallel to the lattice vector c ($z$ axis) and the three bonds forming the basal plane (xy plane) of the tetrahedra, respectively. Moreover, there is one characteristic angle $\alpha$ between these two types of bonds, or alternatively, the angle $\beta$ between each of the two basal plane bonds may be chosen. There is a direct correspondence\cite{Ambacher2002} of these parameters with the lattice parameters $\alat$ and $\clat$ and the internal parameter $u$ of the  wurtzite structure. The respective values are summarized in Tab.\ \ref{tab:binary_phases}.
The situation is more complicated in (Al,Sc)N where the MN$_4$ tetrahedra (we use the notation M for a \emph{metal atom}) in general have a broken symmetry which results in many more parameters that are needed for their characterization, see Fig.\ \ref{fig:geom_def}. All four tetrahedral \mbox{M--N} bonds can have different lengths. With $\ell_c$ we refer to the bond length of the \mbox{M--N} bond which is oriented roughly in c direction but may have a small tilting angle $\gamma$ with respect to the $z$ axis. The \mbox{M--N} bonds involving the three N atoms in the basal plane in general have three different lengths $\ell_{ab,i}$ (i=1,2,3) and three different angles $\delta_i$ measuring the tilt with respect to the $xy$ plane. The three bond angles $\alpha_i$ differ as well, the same applies to the three $\beta_i$.

\begin{figure}[]
\begin{center}
\setlength{\unitlength}{1mm}
\begin{picture}(85,133)(0,0)
\put(0,52){\includegraphics[width=\columnwidth]{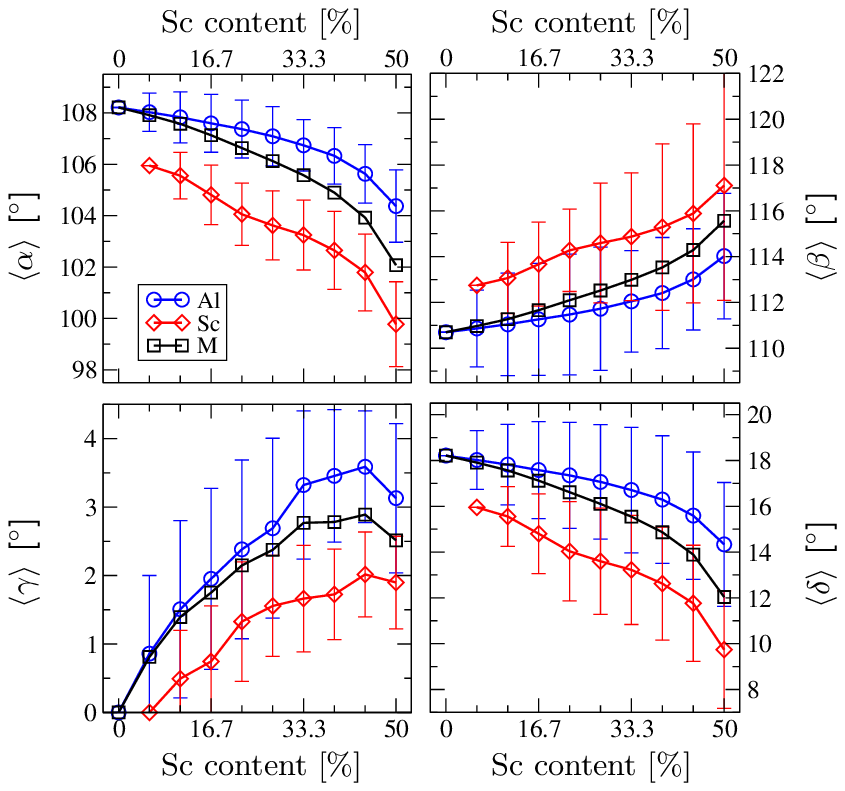}}
\put(0,0) {\includegraphics[width=\columnwidth]{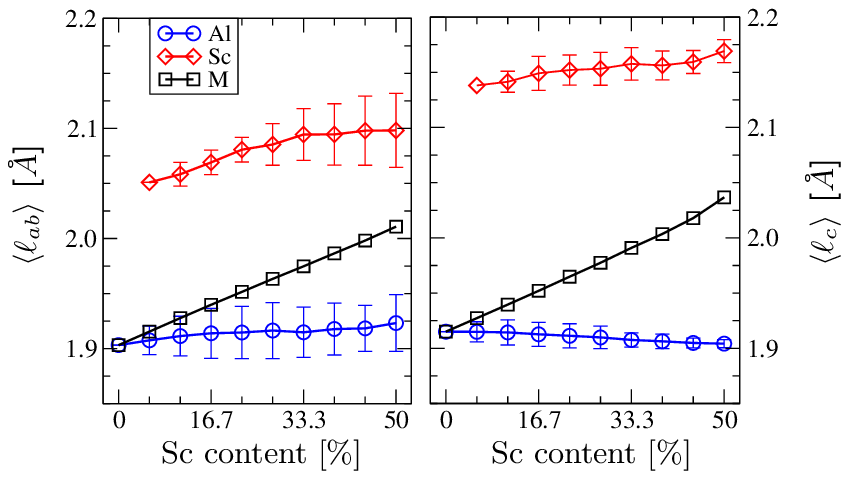}}
\end{picture}
\caption{
Bond lengths and bond angles averaged over the set of structurally relaxed low-energy supercell models as a function of the Sc content. Values from AlN$_4$ tetrahedra (blue circles) and ScN$_4$ tetrahedra (red diamonds) are shown separately and the black symbols mark their weighted average at given Sc content $x$. Error bars indicate $\pm$ one standard deviation from the average value.
\label{fig:average_geom}}
\end{center}
\end{figure} 

We have conducted a statistical analysis in order to derive a correspondence between the lattice parameters and these bond lengths and bond angles that in general vary for all considered atomic bonds. Therefore we have averaged each of the parameters $\ell_{ab}$, $\ell_{c}$, $\alpha$, $\beta$, $\gamma$ and $\delta$ over all the Al--N and Sc--N bonds in all the structurally relaxed supercells of our dataset at each given Sc content. The respective results are summarized in Fig.\ \ref{fig:average_geom}. The histograms for Al--N and Sc--N bonds have been evaluated separately and the blue circles and red diamonds give the respective mean values. Error bars of $\pm$ one standard deviation indicate the spread of the distributions. Finally, the averaged values weighted by the respective Al:Sc ratios are shown as black squares.

The average Al--N bond lengths $\langle\ell_{ab}\rangle_{\rm{Al}}$ and $\langle\ell_c\rangle_{\rm{Al}}$ are found to be very close to the values of bulk wurtzite AlN, c.f.\ Table \ref{tab:binary_phases}. Moreover, both lengths vary only marginally with the Sc content. By contrast, the values for the Sc--N bond lengths are found to be substantially smaller than the values of cubic ScN and hexagonal ScN and they gradually grow with increasing number of Sc atoms in the supercell.
For the bond angles, the situation is different. While $\langle\alpha\rangle_{\rm{Sc}}$ is a few degrees smaller than $\langle\alpha\rangle_{\rm{Al}}$, both decrease monotonously and with a similar slope as a function of the Sc content. The disorder and symmetry breaking introduced by the Sc atoms leads to a tilting of the MN$_4$ tetrahedra as reflected in the increase of $\langle\gamma\rangle$ for both M = Al and Sc. Here, the latter is less affected than the former. The observed decrease in $\langle\delta\rangle$ is directly connected with the decrease in $\langle\alpha\rangle$. Since $\langle\gamma\rangle$ is small, the relation $\langle\delta\rangle \simeq \langle\alpha\rangle - 90^\circ$ holds to a large extent.

The lattice parameter $\alat$ can be obtained by averaging over the projection of the M--N bonds onto the $xy$ plane 
\begin{eqnarray}
\label{eq:a_approx}
     \langle a\rangle&=&\sqrt{3}\,\langle{\cal P}_{\!xy}\ell_{ab}\rangle
     \simeq\sqrt{3}\;\langle\ell_{ab}\rangle\;\sin\langle\alpha\rangle.
\end{eqnarray}
The averaged projection $\langle{\cal P}_{\!xy}\ell_{ab}\rangle$ grows even faster than the average $\langle\ell_{ab}\rangle$ with increasing Sc content because  $\langle\alpha\rangle$ is decreasing. Both effects add up and give the observed almost linear dependence of $\alat$ on $x$. By contrast, the averaged projection onto the z axis $\langle{\cal P}_{\!z}\ell_{ab}\rangle$ decreases for the same reason and largely compensates the increase of $\langle\ell_c\rangle$ with increasing Sc content. This in combination leads to the observed dependence of lattice parameter $\clat$ on $x$ since
\begin{eqnarray}
\label{eq:c_approx}
\langle c\rangle &=&2\langle{\cal P}_{\!z}\ell_{c}\rangle+2\langle{\cal P}_{\!z}\ell_{ab}\rangle
\nonumber\\
 &\simeq&2\langle\ell_{c}\rangle\cos\langle\gamma\rangle+2\langle\ell_{ab}\rangle\sin\langle\delta\rangle.
\end{eqnarray}
For small values of $\langle\gamma\rangle$ equation (\ref{eq:c_approx}) can be well approximated by setting 
$\sin\langle\delta\rangle\simeq-\cos\langle\alpha\rangle$ and $\cos\langle\gamma\rangle\simeq1$.

\section{Elastic tensor}
\label{sec:Elastic_tensor}

This section introduces our method of evaluating the elastic tensor as a function of Sc content $x$
in Sec.\ \ref{subsec:Evaluation_Cij}, and the respective results are presented in Sec.\ \ref{subsec:resultsCij}. The observed behavior can be well correlated with the change in the interatomic bonds due to an applied strain as shown and discussed in Sec.\ \ref{subsec:origin_softening}.

\subsection{Evaluation of tensor components $C_{\mu\nu}$}
\label{subsec:Evaluation_Cij}

\begin{figure*}[t]
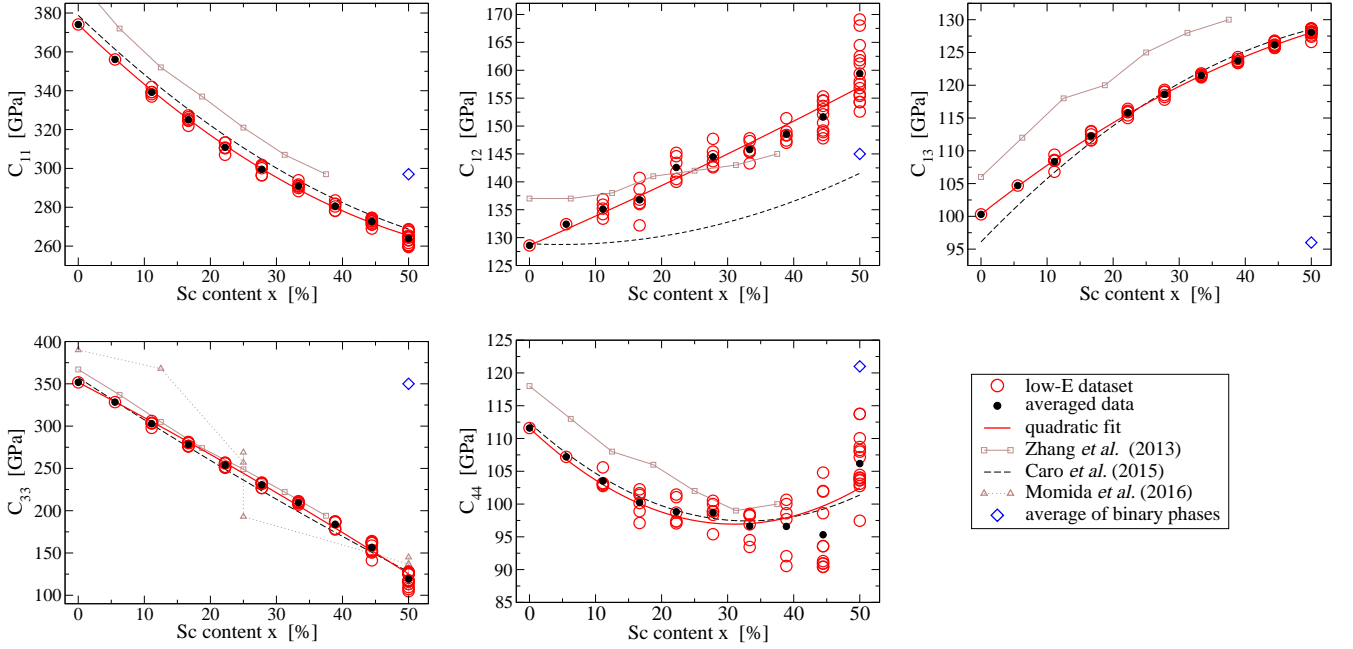

\begin{center}
\setlength{\unitlength}{1mm}
\begin{picture}(175,90)(0,0)
\put(0,45){\includegraphics[width=0.65\columnwidth]{Fig6a_C11.eps}}
\put(60,45){\includegraphics[width=0.65\columnwidth]{Fig6b_C12.eps}}
\put(120,45){\includegraphics[width=0.65\columnwidth]{Fig6c_C13.eps}}
\put(0,0){\includegraphics[width=0.65\columnwidth]{Fig6d_C33.eps}}
\put(60,0){\includegraphics[width=0.65\columnwidth]{Fig6e_C44.eps}}
\put(128,15){\includegraphics[width=0.4\columnwidth]{Fig6f_Cij_legend.eps}}
\end{picture}
\caption{
Symmetrized elastic tensor components $C_{\mu\nu}$ calculated for 36-atom (Al,Sc)N supercells with varying Sc content x. The results of the quadratic fitting are marked as solid lines. Results taken from Refs. [\onlinecite{Zhang2013,Caro2015,Momida2016}] are included for comparison. The arithmetic mean of the elastic properties of binary AlN and ScN is displayed by blue diamond symbols at x=50\%.
\label{fig:Cij}}
\end{center}
\end{figure*}

The tensor of elastic constants is of rank 4, which implies tensor components $C_{ijkl}$ with four cartesian indices. Due to symmetry, it is convenient to use the Voigt notation in order to write the tensor components in matrix form with elements $C_{\mu\nu}$  (and with $C_{\mu\nu}=C_{\nu\mu}$). Here and in the following we use Greek letters for tensor indices in Voigt notation and Latin letters for the cartesian indices.  
The matrix representing the elastic tensor has five independent non-zero components 
for the hexagonal symmetry of the wurtzite crystal. These are
$C_{11}$, $C_{12}$, $C_{13}$, $C_{33}$, and $C_{44}$.  By symmetry $C_{22}=C_{11}$, $C_{23}=C_{13}$, $C_{55}=C_{44}$, and  $C_{66}=(C_{11}-C_{12})/2$. 

The random distribution of Sc atoms on the metal sublattice a priori breaks the hexagonal symmetry for the considered (Al,Sc)N supercell, so that a calculation will yield the full set of 21 non-vanishing independent components $C_{\mu\nu}$. In order to restore the hexagonal symmetry of the elastic tensor as it is observed experimentally for (Al,Sc)N films on the macroscopic level, we make use of the appropriate point group symmetry C$_{\rm 6v}$ (6mm). This group comprises 12 symmetry elements, namely five rotation angles (60$^\circ$, 120$^\circ$, 180$^\circ$, 240$^\circ$, 300$^\circ$), six mirror planes and the identity. The transformation of the elastic tensor (tensor of rank 4) under a symmetry operation $R(\alpha)$ with corresponding transformation matrices $R_{ij}^{(\alpha)}$ is given by
\begin{equation}
C^{(\alpha)}_{ijkl}=\sum_{m=1}^3\sum_{n=1}^3\sum_{p=1}^3\sum_{q=1}^3 
R_{im}^{(\alpha)} R_{jn}^{(\alpha)}R_{kp}^{(\alpha)}R_{lq}^{(\alpha)}C_{mnpq}\,.
\end{equation} 
The symmetrized tensor is obtained as an average with respect to the 12 symmetry elements of the point group symmetry C$_{\rm 6v}$,
\begin{equation}
C^{(sym.)}_{ijkl}=\frac{1}{12}\sum_{\alpha=1}^{12}C^{(\alpha)}_{ijkl}\,.
\end{equation}

We have evaluated the elastic tensor from first-principles stress calculations for all of the selected 36-atom sample structures using the ElaStic package.\cite{Golesorkhtabar2013} Depending on the space group of the crystal, a set of deformation matrices is selected. Stress calculations are carried out for all deformed structures and the computed stresses are fitted as polynomial functions of the applied strains in order to extract the derivatives at zero strain. The knowledge of these derivatives allows for the determination of all independent components of the elastic tensor. In this context, the accuracy of the elastic constants critically depends on the polynomial fit, 
namely the order of the polynomial used and the range of deformations considered. The ElaStic tool allows for a systematic study of the influence of these fitting parameters on the numerical derivatives in order to obtain the most reliable results. We have used third order polynomials and the strain interval [-0.004, 0.004] with 17 equally spaced data points for each deformation.

\subsection{Results: Elastic tensor of (Al,Sc)N}
\label{subsec:resultsCij}

\begin{table}[]
\begin{center}
\begin{tabular}{l c c c c c}
&$C_{11}\,$[GPa] & $C_{12}\,$[GPa] &$C_{13}\,$[GPa] &$C_{33}\,$[GPa] &$C_{44}\,$[GPa]\\
\hline
w-AlN	     & 374 & 129 & 101& 351 & 112 \\
h-ScN	     & 218 & 160 & 89 & 346 & 132 \\
average      & 296 & 145 & 95 & 349 & 122 \\
\end{tabular}
\caption{Elastic tensor components computed for wurtzite AlN and hexagonal ScN. The third line gives the arithmetic mean of the values of w-AlN and h-ScN. 
\label{tab:comp_Cij_AlN_ScN}}
\end{center}
\end{table}

The results for the symmetrized elastic tensor are compiled in Fig.\ \ref{fig:Cij}, where they are also compared with other DFT results from literature.\cite{Zhang2013,Caro2015,Momida2016}
The elastic tensor components vary for the individual structure models at a given Sc content $x$. However, their averaged values (black circles in Fig.\ \ref{fig:Cij}) change monotonously as a function of $x$. As for the lattice parameters we have fitted a quadratic function to these data with the  constraint that the function at $x=0$ has the value of the AlN parameter. The results of the fit (solid red lines in Fig.\ \ref{fig:Cij}) are given by 
\begin{eqnarray}
C_{11}(x)&=&374.1 \left(1 - 0.882\, x + 0.602\, x^2 \right) {\rm GPa} ,\\
C_{12}(x)&=&128.6 \left(1 + 0.400\, x - 0.082\, x^2 \right) {\rm GPa} ,\\
C_{13}(x)&=&100.3 \left(1 + 0.793\, x - 0.481\, x^2 \right) {\rm GPa} ,\\
C_{33}(x)&=&351.7 \left(1 - 1.160\, x - 0.256\, x^2 \right) {\rm GPa} ,\\
C_{44}(x)&=&111.6 \left(1 - 0.848\, x + 1.369\, x^2 \right) {\rm GPa} .
\end{eqnarray}
The quadratic fitting works very well for all five tensor components. Note however, that there is a small modulation in the $C_{12}$ data around the fitted, almost linear curve which cannot be captured by the quadratic fitting ansatz. Our results agree qualitatively with the values obtained by the other theory groups except from $C_{12}$ for which we predict a considerably stronger increase with increasing Sc content. 

Blue diamond symbols in Fig.\ \ref{fig:Cij} mark the arithmetic mean of the tensor components of pure wurtzite AlN and hexagonal ScN (cf.\ Tab.\ \ref{tab:comp_Cij_AlN_ScN}) at x=50\%.
The comparison of these estimates with our data allows us to distinguish two cases. On the one hand, the change of the elastic tensor components $C_{11}$ and $C_{12}$ with increasing Sc content can be fairly well approximated by the interpolation between the two pure phases. On the other hand, this does not hold for $C_{13}$, $C_{33}$, and $C_{44}$.

Although $C_{33}$ has roughly the same value for the pure components (cf.\ Tab.\ \ref{tab:comp_Cij_AlN_ScN}) there is a considerable softening for the mixed crystal. The behavior of $C_{13}$ as well cannot be inferred from an interpolation between the pure phases which would predict a decrease with growing Sc content instead of the observed increase. Finally, $C_{44}$ is also found to soften; it starts to increase at a large Sc content of $\simeq50\%$ which is directly correlated with the significant nonlinear variation of the lattice parameter $\clat$ in this range.

\subsection{Microscopic origin of $C_{33}$ softening}
\label{subsec:origin_softening}

The elastic constants of other mixed wurtzite-type nitrides, namely (Al,Ga)N, (Al,In)N, or (In,Ga)N,  are found to depend linearly on composition.\cite{Lepkowski2015} Deviations from this Vegard's rule  behavior are small and typically of the order of a few percent only. By contrast, (Al,Sc)N apparently  does not follow this trend.

\begin{figure}[b]
\begin{center}
\setlength{\unitlength}{1mm}
\begin{picture}(85,38)(0,0)
\put(1.5,3){\includegraphics[width=0.95\columnwidth]{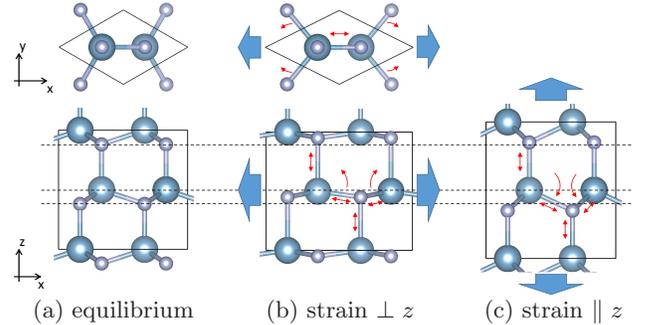}}
\put(4,0){(a) equilibrium}
\put(35,0){(b) strain $\perp z$}
\put(64,0){(c) strain $\parallel z$}
\end{picture}
\caption{Sketch of the average atomic configuration (a) in equilibrium, and under an  applied uniaxial strain (b) within the $xy$ plane and (c) parallel to the $z$ axis. Dashed lines serve as a guide for the eye and mark the equilibrium positions. Little red arrows highlight the directions of the changes in bond lengths and angles under strain. All relative changes are largely magnified for better visibility. 
\label{fig:sketch_strain}}
\end{center}
\end{figure}

The qualitative different dependencies on composition of the C$_{\mu\nu}$ of (Al,Sc)N can be seen as analogous to those of the lattice parameters $\alat$ and $\clat$ on the Al/Sc ratio. Therefore, we correlate the dependence of the elastic tensor components on the Al/Sc-ratio with the microscopic atomic structure. We have analyzed the distribution of bond lengths and angles for our supercell models when a strain is applied either in $z$ direction ($\varepsilon_\parallel$) or in a direction within the $xy$ plane ($\varepsilon_\perp$). The atoms react to the applied strain and the averaged atomic configuration is modified with respect to the equilibrium structure, as sketched in Fig.~\ref{fig:sketch_strain}.
For this comparison we have chosen an uniaxial strain of $\varepsilon=0.004$ which is the maximum applied strain in our calculation of elastic tensors. Hence we have strained the crystal in the respective direction accordingly while keeping the dimensions in the other two directions fixed at their equilibrium values. For the case of an applied strain perpendicular to the $z$ direction we have considered both the $x$ and $y$ directions. Their averages will be discussed in the following. This procedure corresponds once more to averaging over the symmetry equivalent supercell realizations.   

\begin{figure}[]
\begin{center}
\includegraphics[width=\columnwidth]{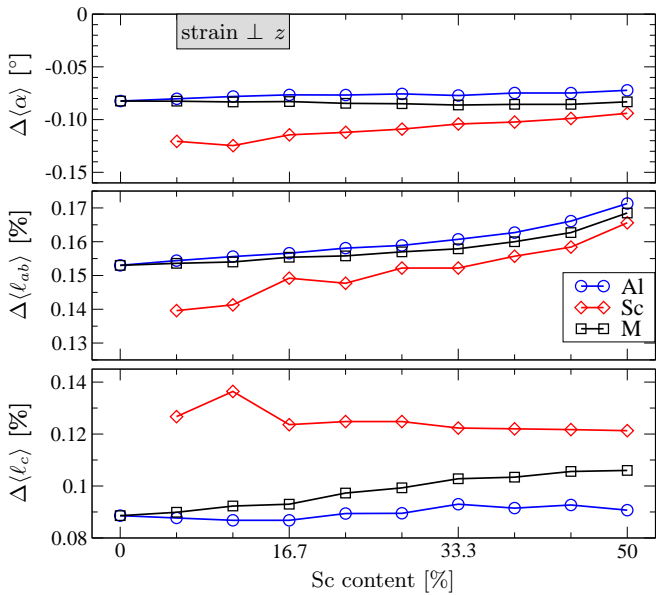}
\caption{Averaged change in bond lengths and bond angles for a strain $\varepsilon_{\perp}=0.004$ applied perpendicular to the $z$ direction. The black square symbols display the weighted average of the results for Al--N (blue circles) and Sc--N (red diamonds) at the respective Sc content $x$. 
\label{fig:strainXY}}
\end{center}
\end{figure} 

The results for the change in the average bond lengths $\langle\ell_{ab}\rangle$ and $\langle\ell_c\rangle$ and bond angle $\langle\alpha\rangle$ due to an applied strain perpendicular to the $z$ direction are visualized in Fig.\ \ref{fig:strainXY}. All quantities are apparently proportional to the Sc content within a wide range of $x$. Moreover, the average elongation $\langle\ell_c\rangle$ of the bonds in $z$ direction and the decrease in bond angle  $\langle\alpha\rangle$ are nearly independent of the Sc content for both, Al--N and Sc--N bonds. 
As the Al--N bonds are stiffer than the Sc--N bonds, the admixture of Sc leads to the observed softening of $C_{11}$ and $C_{12}$ which roughly follows the linear interpolation between the two binary compounds. The nonlinearity in both tensor components can be traced back to the response of $\langle\ell_{ab}\rangle$ (Fig.\ \ref{fig:strainXY}, middle panel). The three bonds forming the basal plane of the MN$_4$ tetrahedra are forced to take up more of the applied strain the more the bond angle $\alpha$ decreases with increasing Sc content.

\begin{figure}[]
\begin{center}
\includegraphics[width=\columnwidth]{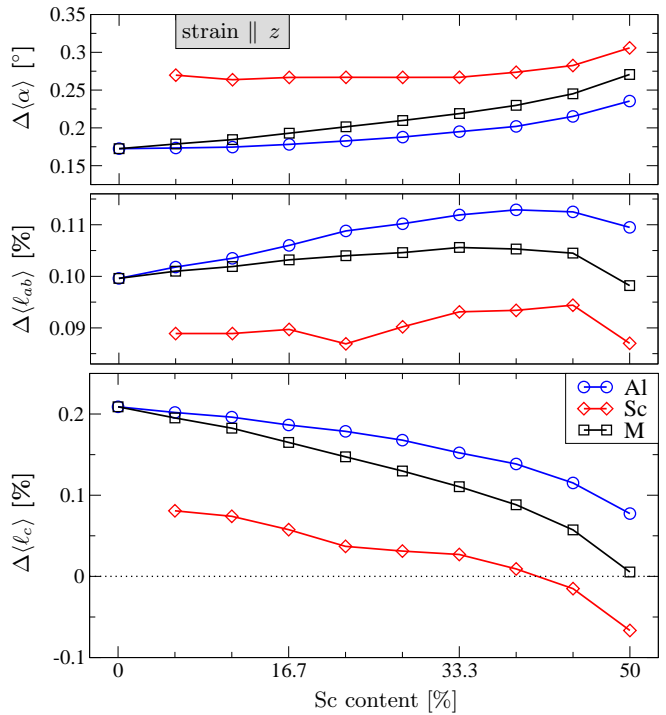}
\caption{Averaged change in bond lengths and bond angles for a strain $\varepsilon_{\parallel}=0.004$ applied in $z$ direction. The black square symbols display the weighted average of the results for Al--N (blue circles) and Sc--N (red diamonds) at the respective Sc content x. 
\label{fig:strainZ}}
\end{center}
\end{figure}

The situation is qualitatively different for an applied strain in $z$ direction as visualized in Fig.\ \ref{fig:strainZ}. Here, the interplay of changing Al--N and Sc--N bond lengths and bond angles leads to an overall decreasing strain on the bonds in $z$ direction, which is reflected in a decrease of $\langle\ell_{c}\rangle$ as a function of $x$. Opposed to that, the response of $\langle\ell_{ab}\rangle$ is almost independent of x when averaged over all metal atoms. The strong decrease of $\Delta\langle\ell_{c}\rangle_M$ with $x$ is directly related to the observed softening of $C_{33}$ and is accomplished by a considerable change in the average bonding angle $\langle\alpha\rangle_M$. In other words, most of the applied strain in $z$ direction is reflected in the increase of the projection of the basal plane bonds onto the $z$ axis,
\begin{eqnarray}
	\langle{\cal P}_{z}\ell_{ab}\rangle\simeq\langle\ell_{ab}\rangle\sin\langle\alpha\rangle .
\end{eqnarray}
This length measures the average distance between the M- and N-planes. 

\section{Piezoelectric tensor}
\label{sec:Piezo_tensor}

In the following, we describe the method to compute the $x$-dependent piezoelectric tensor in 
Sec.\ \ref{subsec:Evaluation_eij}. Results are presented in Sec.\ \ref{subsec:resultseij}. They are analyzed and traced back to their microscopic origin in Secs.\ 
\ref{subsec:Origin_piezo} and \ref{subsec:competition_bonding}.

\subsection{Evaluation of tensor components $e_{i\mu}$}
\label{subsec:Evaluation_eij}

The piezoelectric tensor is of rank 3 with tensor components $e_{ijk}$. Frequently, the second and third Cartesian indices are merged into one index in the Voigt notation, so that the tensor components can be written in matrix form with elements $e_{i\mu}$. Given the hexagonal symmetry of the wurtzite structure, this matrix has three independent non-zero coefficients. These are $e_{15}$, $e_{31}$, and $e_{33}$; by symmetry $e_{32}=e_{31}$ and $e_{25}=e_{15}$. The random distribution of Sc atoms on the metal sublattice breaks the symmetry for the considered (Al,Sc)N supercell, so that there will be the full set of 18 independent components.
Corresponding to our workflow for the calculation of elastic constants, we make use of the point group symmetry C$_{\rm 6v}$ (6mm) in order to restore the hexagonal symmetry of the piezoelectric tensor as it is observed experimentally on the macroscopic level. The symmetry averaged tensor components 
are obtained from
\begin{equation}
e^{(sym.)}_{ijk}=\frac{1}{12}\sum_{\alpha=1}^{12}e^{(\alpha)}_{ijk}
\end{equation}
with
\begin{equation}
e^{(\alpha)}_{ijk}=\sum_{m=1}^3\sum_{n=1}^3\sum_{p=1}^3 
R_{im}^{(\alpha)} R_{jn}^{(\alpha)}R_{kp}^{(\alpha)}e_{mnp},
\end{equation}
using the 12 symmetry elements $R^{(\alpha)}$ of point group C$_{\rm 6v}$ with corresponding transformation matrices $R_{ij}^{(\alpha)}$.

For the determination of the piezoelectric tensors we have adapted and extended the workflow as implemented in the ElaStic tool. Following the \emph{modern theory of polarization} \cite{Vanderbilt2000, Resta2007} the piezoelectric response is related to the dependence of the Berry phase $\phi$ on the elastic strain,
\begin{equation}
e_{ijk} = \frac{1}{2\pi}\frac{e}{\Omega}\sum_{\alpha=1}^3\frac{\phi_\alpha}{\varepsilon_{jk}\;}r_{\alpha, i}.
\end{equation} 
Here, $\varepsilon_{jk}$ is a strain tensor component, $r_{\alpha, i}$ the $i$-th component of one of the three (primitive) lattice vectors $\vec{r}_{\alpha}$, $\Omega=\vec{r}_1\cdot(\vec{r}_{2}\times\vec{r}_3)$ is the unit-cell volume, and $e$ is the electron charge. 
The Berry phase is computed for the three primitive reciprocal lattice vectors $\vec{g}_\alpha$ (corresponding to the real-space lattice vectors $\vec{r}_\alpha$),
\begin{equation}
	\phi_\alpha = \frac{1}{\Omega_{\rm{BZ}}}{\rm{Im}}\sum_{n \rm (occ.)}\int_{\rm{BZ}}d^3k\left\langle u_{n\vec{k}} \left|
	\vec{g}_\alpha\cdot\vec{\nabla}_{\vec{k}}\right|u_{n\vec{k}}\right\rangle.
\end{equation}
Here $\Omega_{\rm{BZ}}$ is the volume of the first Brillouin zone and the $u_{n\vec{k}}$ are the Bloch states. The sum includes all occupied bands.

We use the same set of deformation matrices and strained deformed structure models as in Sec.\ \ref{sec:Elastic_tensor} for the analysis of elasticity. Calculations of the Berry phase are carried out using the implementation in QE for each deformed structure. 
We have used 5, 5, and 11 discrete k-points for the integration along the three reciprocal lattice directions. Subsequently, the data are fitted as third order polynomial functions of the applied strains in order to extract the derivatives at zero strain. The knowledge of these derivatives allows for the determination of all independent components of the piezoelectric tensor. 

The piezoelectric tensor coefficients are commonly discussed by dividing them into two parts.\cite{Bernardini1997} (i) The first part captures the change in polarisation due to a straining of the lattice. This so-called \emph{clamped-ion term} represents the effect of external macroscopic strain on the electronic structure. It is computed without a relaxation of interatomic forces in the strained  structure models. (ii) The second part to the piezoelectric tensor coefficients then reflects the presence of internal strain. It explicitly involves the piezoelectric response with respect to the change in internal structure parameters by displacements of atoms induced by the strain.

\subsection{Results: Piezoelectric tensor of (Al,Sc)N}
\label{subsec:resultseij}

\begin{figure}[]
\begin{center}
\includegraphics[width=\columnwidth]{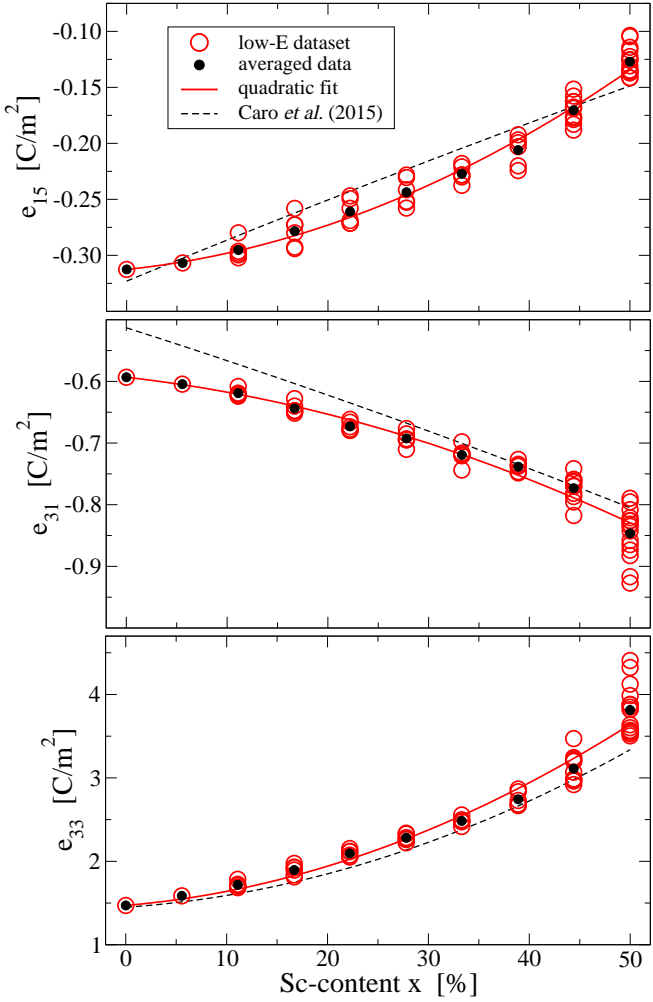}
\caption{
Symmetrized piezoelectric tensor components calculated for 36-atom (Al,Sc)N supercells with varying Sc content x. The results of the quadratic fitting are shown as solid lines. Data taken from Ref.\ [\onlinecite{Caro2015}] are shown as dashed lines for comparison.
\label{fig:eij}}
\end{center}
\end{figure} 

We have evaluated the full set of tensor components for the subset of low-energy sample structures.
The results for the symmetrized piezoelectric tensor are presented in Fig.\ \ref{fig:eij}, and  compared there with other DFT results from literature.\cite{Caro2015}
The individual structure models yield varying tensor components, like what was observed for the elastic tensor in Sec.\ \ref{subsec:resultsCij}. Nevertheless, the averaged values at each given Sc content (black circles in Fig.\ \ref{fig:eij}) change monotonously as a function of $x$. We have fitted a quadratic function to the data thereby constraining the function at $x=0$ to the AlN parameters. The results of the fit (solid red lines in Fig.\ \ref{fig:eij}) are given by 
\begin{eqnarray}
e_{15}(x) &=& -0.313 \left(1 - 0.296\, x - 1.687\, x^2 \right) {\rm C/m}^2, \\
e_{31}(x) &=& -0.593 \left(1 + 0.311\, x + 0.971\, x^2 \right) {\rm C/m}^2 ,\\
e_{33}(x) &=&  1.471 \left(1 + 0.699\, x + 4.504\, x^2 \right) {\rm C/m}^2.
\end{eqnarray}
All three components vary significantly as a function of $x$. While $e_{15}$ decreases by $\sim57\%$ in magnitude when x is increased from 0 up to 50$\%$, $e_{31}$ increases by $\sim40\%$ in the same x range.
Most notably $e_{33}$ increases by $\sim150\%$ when comparing (Al,Sc)N with 50$\%$ Sc with pure AlN.

\begin{figure}[]
\begin{center}
\includegraphics[width=\columnwidth,draft=false]{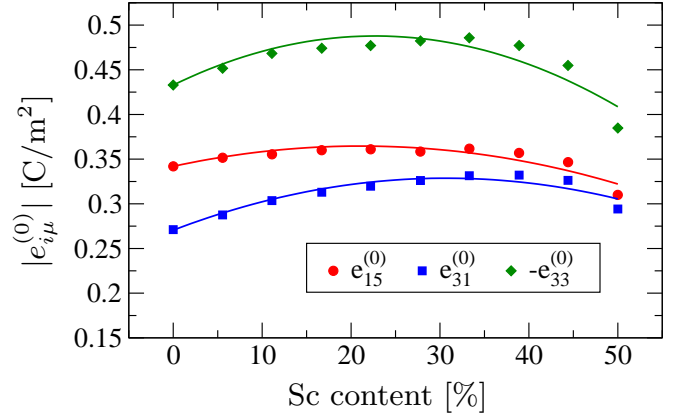}
\caption{ 
Clamped-ion contributions $e_{i\mu}^{(0)}$ to the respective piezoelectric coefficients. Symbols mark the symmetry averaged values for each specific Sc concentration. The result of the quadratic fitting is shown as solid lines. Note that $e_{33}^{(0)}$ is opposite in sign compared with $e_{15}^{(0)}$ and $e_{31}^{(0)}$.
\label{fig:eij_clamped_ion}}
\end{center}
\end{figure}

For further analysis we single out the clamped-ion terms $e_{15}^{(0)}$, $e_{31}^{(0)}$, and $e_{33}^{(0)}$ which are plotted in Fig.\ \ref{fig:eij_clamped_ion}. They do not contribute strongly to the large variations of the full piezoelectric coefficients as can be seen by direct comparison with Fig.\ \ref{fig:eij}.

\subsection{Microscopic origin of significant non-linear increase in $e_{33}$}
\label{subsec:Origin_piezo}

\begin{figure}[]
\begin{center}
\setlength{\unitlength}{1mm}
\includegraphics[width=\columnwidth]{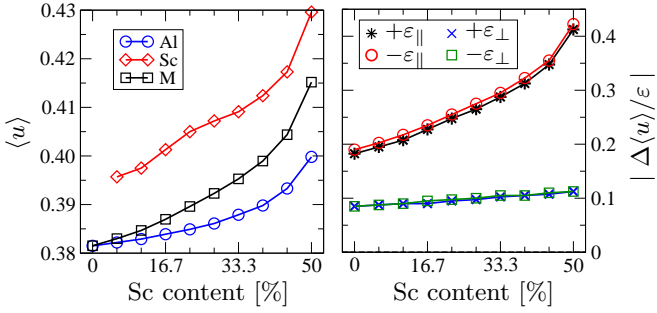}
\caption{Left: Average internal parameter $\langle u \rangle$ as a function of Sc content. Values from AlN$_4$ and ScN$_4$ tetrahedra are averaged separately and the black symbols show their weighted average. Right: Change in the average parameter $\langle u \rangle$ due to uniaxial strain of $\pm 0.004$ in z direction ($\parallel$) or within the xy plane ($\perp$).
\label{fig:u_of_x}}
\end{center}
\end{figure}

As described in the previous sections, the set of supercell models of our study maps to a wurtzite crystal if the results are statistically averaged according to the hexagonal C$_{\rm 6v}$ point group symmetry. Therefore it is possible to define an averaged $u$ parameter as
\begin{equation}
\label{eq:def:u}
	\langle u \rangle = \frac{1}{2}\frac{\langle{\cal P}_{z}\ell_{c}\rangle}{\langle{\cal P}_{z}\ell_{ab}\rangle + \langle{\cal P}_{z}\ell_{c}\rangle},
\end{equation}
where $\langle{\cal P}_{z}\ell_{ab}\rangle$  and $\langle{\cal P}_{z}\ell_{c}\rangle$ are the averaged projections onto the z axis of the $\ell_{ab}$ and $\ell_{c}$ bonds, respectively (cf.\ Figs.\ \ref{fig:wurtzite} and \ref{fig:geom_def}). The dependence of $\langle u \rangle$ on the Sc content is shown in the left panel of Fig.\ \ref{fig:u_of_x}. This reflects the $x$ dependence of the piezoelectric coefficient $e_{33}$ and an almost linear relation between $e_{33}$ and $\langle u \rangle$ is found.
However, we need to consider the response of $\langle u \rangle$ with respect to strain in order to establish a more satisfactory correlation with the microscopic parameters which captures both, the variations in $e_{33}$ and $e_{31}$.

The piezoelectric tensor coefficient $e_{33}$ of wurtzite crystals is commonly discussed by dividing it into the following two parts,\cite{Bernardini1997}
\begin{eqnarray}
\label{eq:decomposition_eij_general}
	e_{33} &=&\clat\,\frac{\partial P_z}{\partial\clat}+\frac{4e{\cal Z}^*}{\sqrt{3}\alat^2}\frac{du}{d\varepsilon_\parallel}.
\end{eqnarray}
Here ${\cal Z}^*$ is the dynamical Born charge in units of the electronic charge $e$ and $\varepsilon_\parallel$ is the applied strain in $z$ direction. 
The clamped-ion term $e_{33}^{(0)}=\clat\;\partial P_z/\partial\clat$ captures the change in polarisation $P_z$ in $z$ direction due to a macroscopic strain on the lattice.  The second term in Eq.\ (\ref{eq:decomposition_eij_general}) reflects the presence of internal strain and explicitly involves the derivative of the internal wurtzite structure parameter $u$ with respect to strain. The Born dynamical charge itself is defined via the partial derivative of the piezoelectric polarisation  with respect to $u$,\cite{Bernardini1997}
\begin{eqnarray}
\label{def:Zstar}
	{\cal Z}^*&=&\frac{\sqrt{3}\alat^2}{4e}\frac{\partial P_z}{\partial u}.
\end{eqnarray}
Note than an equation analogous to Eq. (\ref{eq:decomposition_eij_general}) holds for $e_{31}$ which then involves the derivative of $u$ with respect to a strain $\varepsilon_\perp$ applied in the $xy$ plane.

We postulate that Eq.\ (\ref{eq:decomposition_eij_general}) also holds for the case of disordered (Al,Sc)N when the internal parameter $u$ of the wurtzite crystal is replaced by the average $\langle u \rangle$,  Eq.\ (\ref{eq:def:u}). When the derivative in the second term is replaced by a finite difference and we make use of Eq.\ (\ref{def:Zstar}), we obtain 
\begin{eqnarray}
\label{eq:e31:decom}
	e_{31}(x) &=& e_{31}^{(0)}(x)+\frac{\partial P_z}{\partial \langle u\rangle}\frac{\Delta\langle u\rangle}{\varepsilon_\perp},
	\\
\label{eq:e33:decom}
	e_{33}(x) &=& e_{33}^{(0)}(x)+\frac{\partial P_z}{\partial \langle u\rangle}\frac{\Delta\langle u\rangle}{\varepsilon_\parallel}.
\end{eqnarray}
Here $\Delta\langle u\rangle$ is the change in the average $\langle u\rangle$ when a uniaxial strain 
$\varepsilon_{\parallel}$ or $\varepsilon_{\perp}$ is applied. The quantity $\Delta\langle u \rangle$ at a strain of $\pm 0.004$ is plotted for the four different cases in the right panel of Fig.\ \ref{fig:u_of_x}.
A positive strain in $z$ direction ($+\varepsilon_\parallel$) leads to a decrease of $\langle u \rangle$ while a positive strain applied in the $xy$ plane ($+\varepsilon_\perp$) yields an increase of the latter. This behavior is reversed for negative strain. The counteracting response and the different magnitude of $\Delta\langle u \rangle$  for the two cases $\varepsilon_\parallel$ and $\varepsilon_\perp$ are reflected in the opposite signs of $e_{33}$ and $e_{31}$ and their magnitudes. 

\begin{figure}[]
\begin{center}
\includegraphics[width=0.6\columnwidth]{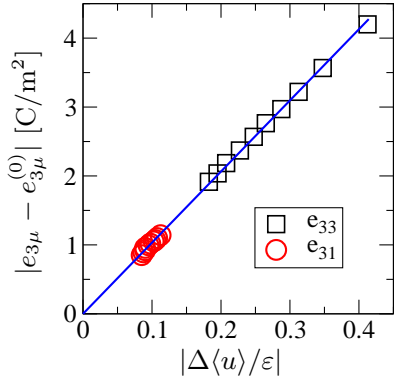}
\caption{Correlation of the piezoelectric coefficients $e_{31}$ and $e_{33}$ with the change in the $u$ parameter due to an uniaxial strain $\varepsilon_\perp$ and $\varepsilon_\parallel$, respectively. A linear fit to the data is shown as blue line.
\label{fig:eij_correlation}}
\end{center}
\end{figure}

Figure \ref{fig:eij_correlation} plots the second terms of Eqs. (\ref{eq:e31:decom}) and (\ref{eq:e33:decom}), i.e. the differences $e_{31}-e_{31}^{(0)}$ and $e_{33}-e_{33}^{(0)}$, as a function of ${\Delta\langle u\rangle}/{\varepsilon}$. A linear correlation is demonstrated which holds for both datasets. As a consequence thereof, the derivative ${\partial P_z}/{\partial \langle u\rangle}$ does not vary significantly as a function of the Sc content and is constant
to leading order. 

In summary, the non-linear increase of $e_{33}$ has its origin essentially in the internal structural distortions induced by straining the crystal in z-direction. 
The local structural sensitivity to the applied strain increases when the Sc content is raised, which is reflected in the dependence of $\langle u \rangle$ on $x$. A microscopic reason for this behavior is given in the following section. Hereby we extend and consolidate the seminal analysis of Ref.\ [\onlinecite{Tasnadi2010}].

\begin{table*}[]
\begin{center}
\begin{tabular}{l l c c c c c}
\hline\hline
Reference&Method&$C_{11}$ [GPa]& $C_{12}$ [GPa]&$C_{13}$ [GPa]&$C_{33}$ [GPa]&$C_{44}$ [GPa]\\
\hline
Kazan \emph{et al.}\cite{Kazan2007}       & Experiment, single crystal, BLS & 394 & 134 & 95  & 402 & 121\\
Sotnikov \emph{et al.}\cite{Sotnikov2010} & Experiment, single crystal, BAW & $402.5\pm0.5$ & $135.6\pm0.5$ & $101\pm2$ & $387.6\pm1$ & $122.9\pm0.5$\\
McNeil \emph{et al.}\cite{McNeil1993} 	  & Experiment, single crystal, BLS & $411\pm10$ & $149\pm10$ & $99\pm4$ & $389\pm10$ & $125\pm5$\\
Deger \emph{et al.}\cite{Deger1998}	      & Experiment, thin film, SAW      & 410 & 140 & 100 & 390 & 120\\
Tsubouchi  \emph{et al.}\cite{Tsubouchi1981}& Experiment, thin film, SAW    & 345 & 125 & 120 & 395 & 118\\
Kurz \emph{et al.}\cite{Kurz2019}         & Experiment, thin film, SAW      & $404\pm3$ & --- & $103\pm15$ & $375\pm13$ & $124\pm2$ \\
Carlotti \emph{et al.}\cite{Carlotti2017} & Experiment, thin film, BLS      & $392\pm8$ & ---  & $106\pm15$  & $385\pm10$ & $112\pm2$\\
\hline
this work	                  & DFT, PWPP (QE), PBE, stress-strain & 376 & 129 & 102 & 353 & 111\\
de Jong \emph{et al.}\cite{deJong2015a} & DFT, PWPP (VASP), PBE, stress-strain & 375 & 130 & 98  & 353 & 113\\
Zhang \emph{et al.}\cite{Zhang2013}	  & DFT, PWPP (VASP), PBE, stress-strain & 397 & 137 & 106 & 367 & 118\\
Caro \emph{et al.}\cite{Caro2015}	  & DFT, PWPP (VASP), PBE, stress-strain & 410 & 142 & 110 & 385 & 123\\
Wrigth \emph{et al.}\cite{Wright1997}  & DFT, PWPP, LDA, energy-strain&	396 & 137 & 108 & 373 & 116\\
Kim \emph{et al.}\cite{Kim1996} 	  & DFT, FP-LMTO, LDA, energy-strain& 398 & 140 & 127 & 382 & 96\\
\hline\hline
\end{tabular}
\caption{Comparison of experimentally measured and theoretically predicted elasticity tensor components of AlN taken from literature.
The second column contains information on the employed experimental and numerical methods used (see text for explanation and discussion).
\label{tab:Lit_Cij_AlN}}
\end{center}
\end{table*}

\subsection{Microscopic reason for variation of internal displacement parameter $\langle u \rangle$}
\label{subsec:competition_bonding}

There is an important difference between the group-IIIA simple-metal element Al (or Ga and In) and the group IIIB transition-metal element Sc (or Y and La) in their metal-nitride compounds. On the one hand, for Al the chemical nearest-neighbor bonds to 2s and 2p valence-electron orbitals of N atoms are formed by Al 3s and 3p orbitals. This results in the sp$^3$ hybridization and the tetrahedral coordination [AlN$_4$] in the hexagonal wurtzite structure of AlN. 
On the other hand, for Sc the bonds to N are formed by Sc 3d and 4s orbitals, which leads to the octahedral coordination [ScN$_6$] in the cubic rocksalt structure of ScN. 

Alloying Al and Sc in their nitrides leads to an energetic competition between the sp--sp character of \mbox{Al--N} bonds of tetrahedrally coordinated Al atoms and the \mbox{sd--sp} character of Sc--N bonds of preferential octahedrally coordinated Sc atoms. For (Al,Sc)N alloys with a Sc content $<50\%$ the tetrahedral coordination of the wurtzite structure is energetically favored. As a consequence,
the Sc atoms occupy tetrahedral sites instead of their favored octahedral sites in these wurtzite-type alloys. To avoid this site dilemma, Sc atoms are displaced more than Al atoms from the regular tetrahedral positions. 

In the wurtzite structure, there are connections from a given tetrahedral site to three neighboring octahedral sites and to another neighboring tetrahedral site through the four triangular faces of the tetrahedron. An isolated single Sc atom at a tetrahedral site of the hexagonal N sublattice would be accommodated by a displacement to one of the three neighboring octahedral sites. However, shifting a Sc atom in the (Al,Sc)N nitride with fully occupied nitrogen and metal sublattices to a neighboring octahedral site would lead to a strong repulsion by metal atoms on next neighbor tetrahedral sites. This leaves only one possible way of achieving a better accommodation for a Sc atom: it is displaced along the hexagonal $c$ axis towards the next empty tetrahedral site. However, the displaced Sc atom cannot reach this tetrahedral site, again because of a strong repulsion by next neighbor metal atoms. Therefore, there is a balance of bonds and forces for Sc atoms close to the triangular N face between two connected tetrahedral sites. This approximately triangular Sc coordination has an internal displacement parameter value of $u\approx1/2$ instead of $u\approx3/8$ for the tetrahedral Al coordination. 

In response to such local displacements of the Sc atoms the Al atoms get displaced as well in the relaxed random-alloy structure, but to a lesser extent. Altogether, a compromise between chemical Sc[sd]--N[sp] and Al[sp]--N[sp] bonds is a reason for the gradual raise of $\langle u \rangle$ between the two limiting values of $u$ with increasing Sc content (see left panel of Fig.\ \ref{fig:u_of_x}). Note that in the case of hexagonal ScN, there is no competing energy term that arises from deformed Al[sp]--N[sp] bonds and the Sc atoms are allowed to relax to the trigonal bipyramidal site with $u=0.5$ and triangular coordination in the $xy$ plane.

\begin{table*}[]
\begin{center}
\begin{tabular}{l l c c c}
\hline\hline
Reference & Method & $e_{33}$ [C/m$^2$] & $e_{31}$ [C/m$^2$] & $e_{15}$ [C/m$^2$]\\
\hline
Bu \emph{et al.}\cite{Bu2004}               & Experiment, single crystal, SAW & $1.39\pm0.22$ & $-0.58\pm0.23$ & $-0.29\pm0.06$ \\
Sotnikov \emph{et al.}\cite{Sotnikov2010}   & Experiment, single crystal, BAW & $1.34\pm0.10$ & $-0.60\pm0.20$ & $-0.32\pm0.05$ \\
Tsubouchi  \emph{et al.}\cite{Tsubouchi1985}& Experiment, thin film, SAW      & 1.55 & $-0.58$ & $-0.48$ \\
Kurz \emph{et al.}\cite{Kurz2019}           & Experiment, thin film, SAW      & $1.52\pm0.43$ & $-0.54\pm0.05$ & $-0.30\pm0.22$ \\
\hline
this work                               & DFT, PWPP (QE), PBE   & 1.48 & $-0.58$ & $-0.32$ \\
Bernardini \emph{et al.}\cite{Bernardini1997} & DFT, PWPP, LDA  & 1.46 & $-0.60$ &	--- \\
Caro \emph{et al.}\cite{Caro2015}       & DFT, PWPP (VASP), PBE & 1.45 & $-0.51$ & $-0.32$ \\
Momida \emph{et al.}\cite{Momida2016}   & DFT, PWPP (VASP), PBE & 1.39 & $-0.55$ & $-0.30$ \\
de Jong \emph{et al.}\cite{deJong2015b}	& DFT, PWPP (VASP), PBE & 1.46 & $-0.58$ & $-0.29$ \\
\hline\hline
\end{tabular}
\caption{Comparison of experimentally measured and theoretically predicted piezoelectricity tensor components of AlN taken from literature.  
The second column contains information on the employed experimental and theoretical methods used (see text for explanation and discussion).
\label{tab:Lit_eij_AlN}}
\end{center}
\end{table*}

\begin{table*}[]
\begin{center}
\begin{tabular}{l l c c c c c c c}
\hline\hline
Composition & Reference & $C_{11}$ [GPa]&$C_{13}$ [GPa]&$C_{33}$ [GPa]&$C_{44}$ [GPa] &$e_{33}$ [C/m$^2$]  & $e_{31}$ [C/m$^2$]  & $e_{15}$  [C/m$^2$] \\
\hline
Al$_{0.86}$Sc$_{0.14}$N & Experiment\cite{Kurz2019} & $354\pm4$ & $108\pm7$ & $305\pm15$ & $112\pm3$ & $1.88\pm0.07$ & $-0.54\pm0.09$ & $-0.26\pm0.07$ \\
Al$_{0.86}$Sc$_{0.14}$N & This work (rescaled)			  & 359	& 113 & 312 & 113 & 1.80 & $-0.57$ & $-0.28$ \\
Al$_{0.68}$Sc$_{0.32}$N & Experiment\cite{Kurz2019} & $307\pm3$ & $123\pm5$ & $230\pm5$ & $110\pm2$ & $2.80\pm0.12$ & $-0.69\pm0.14$ & $-0.23\pm0.13$ \\
Al$_{0.68}$Sc$_{0.32}$N & This work (rescaled)			& 315 & 124 & 226 & 108 & 2.56 & $-0.65$ & $-0.22$\\
\hline\hline
\end{tabular}
\caption{Comparison of experimentally measured and theoretically predicted elastic and piezoelectric tensor components of AlScN with $14\%$ and $32\%$ Sc content. The theoretically obtained functional dependence of the $C_{\mu\nu}$ end $e_{i\mu}$ have been rescaled with reference to binary AlN as explained in the text.
\label{tab:Compare_AlScN}}
\end{center}
\end{table*}

\section{Summary}
\label{sec:summary}
We have investigated the electroacoustic properties of (Al,Sc)N crystals with the metastable wurtzite structure. A combinatorial approach was chosen  and a large variety of structure models with varying Sc content was analyzed. Thereby we sampled the different local atomic configurations of metal-sublattice disorder. For the chosen set of model structures (63 in total) we have evaluated the equilibrium lattice parameters and atomic positions, as well as the full elastic and piezoelectric tensors. The functional dependence of these properties on the Al:Sc ratio was obtained by an averaging and fitting procedure. Thereby we obtained a consistent set of material parameters for (Al,Sc)N extracted from a large data basis over the full range of experimentally accessible Sc content $0\le x\le 50\%$ .

Moreover, a statistical analysis of the microscopic structural parameters -- bond lengths and bond angles -- was conducted. The response of these parameters to an applied uniaxial strain was compared with their equilibrium averages. All structure models were strained parallel and perpendicular to the z axis. This analysis relates the observed variation in elastic and piezoelectric tensor components as a function of Sc content to the change in the averaged values of specific geometrical quantities.

The anisotropic evolution of the lattice parameters $\alat$ and $\clat$ with increasing Sc content is a consequence of an interplay of increasing average bond lengths and decreasing average bond angle 
$\langle \alpha \rangle$. 

The elastic softening in $z$ direction (C$_{33}$) is related to the disorder in local atomic configurations induced by the presence of the Sc atoms. Therefore, an applied strain $\varepsilon_\parallel$ is distributed over several of the microscopic degrees of freedom which, on average, leads to a reduced stretching of the bond lengths $\ell_c$ .  

The extraordinary non-linear increase in the piezoelectric tensor component (e$_{33}$) has its origin in the increased sensitivity of the averaged parameter $\langle u \rangle$ to strain, the more Sc is added to the (Al,Sc)N crystal. Although $\langle u \rangle$ itself increases towards the value $0.5$ of nonpolar hexagonal ScN, its response to strain largely increases as a function of $x$.

All the above mentioned effects follow from the (energetic) competition between Al atoms that favor the tetrahedral coordination by N atoms in the wurtzite structure, and the Sc atoms that would prefer octahedral coordination and need to accommodate themselves as well as possible. The incorporation of Sc on the metal sublattice leads to the observed statistical distribution of bond lengths and bond angles that break the rigid wurtzite crystal symmetry on the microscopic level. This in turn adds flexibility to the atomic structure of (Al,Sc)N on how to respond to strain which finally determines the outstanding elastic and piezoelectric properties.

\section{Acknowledgement} 
We thank Agne Zukauskaite and Nicolas Kurz for many valuable discussions.

\appendix

\section{Elastic and piezoelectric tensor of AlN}
This appendix gives an overview on available literature results for the material parameters of AlN. The comparison of experimentally measured and theoretically predicted elasticity tensor components is compiled in Table\ \ref{tab:Lit_Cij_AlN} while the respective comparison of piezoelectricity tensor components is given in Table\ \ref{tab:Lit_eij_AlN}. 

Although binary AlN is a well and frequently studied material, there is quite some scatter in the available measured tensor components. These differences are partially caused by different crystal quality including possible elastic and piezoelectric inhomogeneities of the samples under study. Further scatter is caused by the different measurement accuracy of the various experimental methods used: Brillouin light scattering (BLS), surface acoustic wave (SAW) measurements, and bulk acoustic wave (BAW) techniques. 

Also the DFT calculations show a noticeable spread in the predicted values. On the one hand this is due to the different implementations of DFT, such as plain wave pseudopotential codes (PWPP) or the full-potential linear muffin-tin (FP-LMTO) approach, and the numerical settings used for convergence. On the other hand, the choice of exchange-correlation functional also influences the result. While the local density approximation (LDA) in general tends to underestimate the equilibrium bonding lengths between atoms, the PBE functional in general overestimates them. However, the comparison of the results summarized in Tables \ref{tab:Lit_Cij_AlN} and \ref{tab:Lit_eij_AlN} does not give evidence, that one or the other choice is superior for the computation of elastic and piezoelectric properties of AlN thin films. Another degree of variation arises from the approach taken for the derivation of elastic constants, namely either the energy-strain or the stress-strain approach.\cite{Caro2012}

\section{Comparison with experiments on AlScN}
Recently, Kurz \emph{et al.}\cite{Kurz2019} for the first time determined a full set of the electroacoustic properties for (Al,Sc)N thin films experimentally from the same material source using Rayleigh-type waves in SAW resonators for high Sc concentrations up to 32$\%$. 
The results from Ref.\ \onlinecite{Kurz2019} for Al$_{0.86}$Sc$_{0.14}$N and Al$_{0.68}$Sc$_{0.32}$N are compared with our theoretical predictions in 
Tab.\ \ref{tab:Compare_AlScN}. To this end, we have corrected for the systematic error in DFT arising from the overestimation of bonding by the PBE functional by rescaling the functional dependence of the $C_{\mu\nu}$ and $e_{i\mu}$ by a constant prefactor so as to meet the end points of binary AlN. This procedure yields a satisfactory agreement between the theoretically predicted and the experimentally measured parameter dependence on Sc content.

\bibliography{bibliography_AlScN}

\begin{thebibliography}{43}
\expandafter\ifx\csname natexlab\endcsname\relax\def\natexlab#1{#1}\fi
\expandafter\ifx\csname bibnamefont\endcsname\relax
  \def\bibnamefont#1{#1}\fi
\expandafter\ifx\csname bibfnamefont\endcsname\relax
  \def\bibfnamefont#1{#1}\fi
\expandafter\ifx\csname citenamefont\endcsname\relax
  \def\citenamefont#1{#1}\fi
\expandafter\ifx\csname url\endcsname\relax
  \def\url#1{\texttt{#1}}\fi
\expandafter\ifx\csname urlprefix\endcsname\relax\def\urlprefix{URL }\fi
\providecommand{\bibinfo}[2]{#2}
\providecommand{\eprint}[2][]{\url{#2}}

\bibitem[{\citenamefont{Matloub et~al.}(2011)\citenamefont{Matloub, Artieda,
  Sandu, Milyutin, and Muralt}}]{Matloub2011}
\bibinfo{author}{\bibfnamefont{R.}~\bibnamefont{Matloub}},
  \bibinfo{author}{\bibfnamefont{A.}~\bibnamefont{Artieda}},
  \bibinfo{author}{\bibfnamefont{C.}~\bibnamefont{Sandu}},
  \bibinfo{author}{\bibfnamefont{E.}~\bibnamefont{Milyutin}}, \bibnamefont{and}
  \bibinfo{author}{\bibfnamefont{P.}~\bibnamefont{Muralt}},
  \bibinfo{journal}{Appl. Phys. Lett.} \textbf{\bibinfo{volume}{99}},
  \bibinfo{pages}{092903} (\bibinfo{year}{2011}).

\bibitem[{\citenamefont{Umeda et~al.}(2013)\citenamefont{Umeda, Kawai, Honda,
  Akiyama, Kato, and Fukura}}]{Umeda2013}
\bibinfo{author}{\bibfnamefont{K.}~\bibnamefont{Umeda}},
  \bibinfo{author}{\bibfnamefont{H.}~\bibnamefont{Kawai}},
  \bibinfo{author}{\bibfnamefont{A.}~\bibnamefont{Honda}},
  \bibinfo{author}{\bibfnamefont{M.}~\bibnamefont{Akiyama}},
  \bibinfo{author}{\bibfnamefont{T.}~\bibnamefont{Kato}}, \bibnamefont{and}
  \bibinfo{author}{\bibfnamefont{T.}~\bibnamefont{Fukura}},
  \bibinfo{journal}{in 26th International Conference on Micro Electro
  Mechanical Systems (MEMS) (IEEE, Taipei, 2013)} pp.
  \bibinfo{pages}{582--–589} (\bibinfo{year}{2013}).

\bibitem[{\citenamefont{Konno et~al.}(2014)\citenamefont{Konno, Kadota,
  Kushibiki, Ohashi, Esashi, Yamamoto, and Tanaka}}]{Konno2014}
\bibinfo{author}{\bibfnamefont{A.}~\bibnamefont{Konno}},
  \bibinfo{author}{\bibfnamefont{M.}~\bibnamefont{Kadota}},
  \bibinfo{author}{\bibfnamefont{J.-I.} \bibnamefont{Kushibiki}},
  \bibinfo{author}{\bibfnamefont{Y.}~\bibnamefont{Ohashi}},
  \bibinfo{author}{\bibfnamefont{M.}~\bibnamefont{Esashi}},
  \bibinfo{author}{\bibfnamefont{Y.}~\bibnamefont{Yamamoto}}, \bibnamefont{and}
  \bibinfo{author}{\bibfnamefont{S.}~\bibnamefont{Tanaka}},
  \bibinfo{journal}{IEEE International Ultrasonics Symposium (IUS), (IEEE,
  Chicago, 2014)} pp. \bibinfo{pages}{273–--276} (\bibinfo{year}{2014}).

\bibitem[{\citenamefont{Parsapour et~al.}(2018)\citenamefont{Parsapour,
  Pashchenko, Nicolay, and Muralt}}]{Parsapour2018}
\bibinfo{author}{\bibfnamefont{F.}~\bibnamefont{Parsapour}},
  \bibinfo{author}{\bibfnamefont{V.}~\bibnamefont{Pashchenko}},
  \bibinfo{author}{\bibfnamefont{P.}~\bibnamefont{Nicolay}}, \bibnamefont{and}
  \bibinfo{author}{\bibfnamefont{P.}~\bibnamefont{Muralt}},
  \bibinfo{journal}{IEEE Micro Electro Mechanical Systems (MEMS), (IEEE,
  Belfast, 2018)} pp. \bibinfo{pages}{763–--766} (\bibinfo{year}{2018}).

\bibitem[{\citenamefont{Kurz et~al.}(2019)\citenamefont{Kurz, Ding, Urban, Lu,
  Kirste, Feil, Zukauskaite, and Ambacher}}]{Kurz2019}
\bibinfo{author}{\bibfnamefont{N.}~\bibnamefont{Kurz}},
  \bibinfo{author}{\bibfnamefont{A.}~\bibnamefont{Ding}},
  \bibinfo{author}{\bibfnamefont{D.~F.} \bibnamefont{Urban}},
  \bibinfo{author}{\bibfnamefont{Y.}~\bibnamefont{Lu}},
  \bibinfo{author}{\bibfnamefont{L.}~\bibnamefont{Kirste}},
  \bibinfo{author}{\bibfnamefont{N.~M.} \bibnamefont{Feil}},
  \bibinfo{author}{\bibfnamefont{A.}~\bibnamefont{Zukauskaite}},
  \bibnamefont{and} \bibinfo{author}{\bibfnamefont{O.}~\bibnamefont{Ambacher}},
  \bibinfo{journal}{J. Appl. Phys.} \textbf{\bibinfo{volume}{126}},
  \bibinfo{pages}{075106} (\bibinfo{year}{2019}).

\bibitem[{\citenamefont{Ichihashi et~al.}(2014)\citenamefont{Ichihashi,
  Yanagitani, Suzuki, Takayanagi, and Matsukawa}}]{Ichihashi2014}
\bibinfo{author}{\bibfnamefont{H.}~\bibnamefont{Ichihashi}},
  \bibinfo{author}{\bibfnamefont{T.}~\bibnamefont{Yanagitani}},
  \bibinfo{author}{\bibfnamefont{M.}~\bibnamefont{Suzuki}},
  \bibinfo{author}{\bibfnamefont{S.}~\bibnamefont{Takayanagi}},
  \bibnamefont{and}
  \bibinfo{author}{\bibfnamefont{M.}~\bibnamefont{Matsukawa}},
  \bibinfo{journal}{in IEEE International Ultrasonics Symposium (IUS) (IEEE,
  Chicago, 2014)} pp. \bibinfo{pages}{2521–--2524} (\bibinfo{year}{2014}).

\bibitem[{\citenamefont{Ichihashi et~al.}(2016)\citenamefont{Ichihashi,
  Yanagitani, Suzuki, Takayanagi, Kawabe, Tomita, and
  Matsukawa}}]{Ichihashi2016}
\bibinfo{author}{\bibfnamefont{H.}~\bibnamefont{Ichihashi}},
  \bibinfo{author}{\bibfnamefont{T.}~\bibnamefont{Yanagitani}},
  \bibinfo{author}{\bibfnamefont{M.}~\bibnamefont{Suzuki}},
  \bibinfo{author}{\bibfnamefont{S.}~\bibnamefont{Takayanagi}},
  \bibinfo{author}{\bibfnamefont{M.}~\bibnamefont{Kawabe}},
  \bibinfo{author}{\bibfnamefont{S.}~\bibnamefont{Tomita}}, \bibnamefont{and}
  \bibinfo{author}{\bibfnamefont{M.}~\bibnamefont{Matsukawa}},
  \bibinfo{journal}{IEEE Trans. Ultrason. Ferroelectr. Freq. Control}
  \textbf{\bibinfo{volume}{63}}, \bibinfo{pages}{717} (\bibinfo{year}{2016}).

\bibitem[{\citenamefont{Carlotti et~al.}(2017)\citenamefont{Carlotti, Sadhu,
  and Dumont}}]{Carlotti2017}
\bibinfo{author}{\bibfnamefont{G.}~\bibnamefont{Carlotti}},
  \bibinfo{author}{\bibfnamefont{J.}~\bibnamefont{Sadhu}}, \bibnamefont{and}
  \bibinfo{author}{\bibfnamefont{F.}~\bibnamefont{Dumont}},
  \bibinfo{journal}{in IEEE International Ultrasonics Symposium (IUS) (IEEE,
  Washington, DC, 2017)} pp. \bibinfo{pages}{1--–5} (\bibinfo{year}{2017}).

\bibitem[{\citenamefont{Lu et~al.}(2018)\citenamefont{Lu, Reusch, Kurz, Ding,
  Christoph, Kirste, Lebedev, and Zukauskaite}}]{Lu2018}
\bibinfo{author}{\bibfnamefont{Y.}~\bibnamefont{Lu}},
  \bibinfo{author}{\bibfnamefont{M.}~\bibnamefont{Reusch}},
  \bibinfo{author}{\bibfnamefont{N.}~\bibnamefont{Kurz}},
  \bibinfo{author}{\bibfnamefont{A.}~\bibnamefont{Ding}},
  \bibinfo{author}{\bibfnamefont{T.}~\bibnamefont{Christoph}},
  \bibinfo{author}{\bibfnamefont{L.}~\bibnamefont{Kirste}},
  \bibinfo{author}{\bibfnamefont{V.}~\bibnamefont{Lebedev}}, \bibnamefont{and}
  \bibinfo{author}{\bibfnamefont{A.}~\bibnamefont{Zukauskaite}},
  \bibinfo{journal}{Phys. Status Solidi A} \textbf{\bibinfo{volume}{215}},
  \bibinfo{pages}{1700559} (\bibinfo{year}{2018}).

\bibitem[{\citenamefont{Mertin et~al.}(2018)\citenamefont{Mertin, Heinz,
  Rattunde, Christmann, Dubois, Nicolay, and Muralt}}]{Mertin2018}
\bibinfo{author}{\bibfnamefont{S.}~\bibnamefont{Mertin}},
  \bibinfo{author}{\bibfnamefont{B.}~\bibnamefont{Heinz}},
  \bibinfo{author}{\bibfnamefont{O.}~\bibnamefont{Rattunde}},
  \bibinfo{author}{\bibfnamefont{G.}~\bibnamefont{Christmann}},
  \bibinfo{author}{\bibfnamefont{M.-A.} \bibnamefont{Dubois}},
  \bibinfo{author}{\bibfnamefont{S.}~\bibnamefont{Nicolay}}, \bibnamefont{and}
  \bibinfo{author}{\bibfnamefont{P.}~\bibnamefont{Muralt}},
  \bibinfo{journal}{Surf. Coat. Technol.} \textbf{\bibinfo{volume}{343}},
  \bibinfo{pages}{2} (\bibinfo{year}{2018}).

\bibitem[{\citenamefont{Wingqvist et~al.}(2010)\citenamefont{Wingqvist,
  Tasnadi, Zukauskaite, Birch, Arwin, and Hultman}}]{Wingqvist2010}
\bibinfo{author}{\bibfnamefont{G.}~\bibnamefont{Wingqvist}},
  \bibinfo{author}{\bibfnamefont{F.}~\bibnamefont{Tasnadi}},
  \bibinfo{author}{\bibfnamefont{A.}~\bibnamefont{Zukauskaite}},
  \bibinfo{author}{\bibfnamefont{J.}~\bibnamefont{Birch}},
  \bibinfo{author}{\bibfnamefont{H.}~\bibnamefont{Arwin}}, \bibnamefont{and}
  \bibinfo{author}{\bibfnamefont{L.}~\bibnamefont{Hultman}},
  \bibinfo{journal}{Appl. Phys. Lett.} \textbf{\bibinfo{volume}{97}},
  \bibinfo{pages}{112902} (\bibinfo{year}{2010}).

\bibitem[{\citenamefont{Feil et~al.}(2019)\citenamefont{Feil, Kurz, Urban,
  Altayara, Christian, Ding, Zukauskaite, and Ambacher}}]{Feil2019}
\bibinfo{author}{\bibfnamefont{N.~M.} \bibnamefont{Feil}},
  \bibinfo{author}{\bibfnamefont{N.}~\bibnamefont{Kurz}},
  \bibinfo{author}{\bibfnamefont{D.~F.} \bibnamefont{Urban}},
  \bibinfo{author}{\bibfnamefont{A.}~\bibnamefont{Altayara}},
  \bibinfo{author}{\bibfnamefont{B.}~\bibnamefont{Christian}},
  \bibinfo{author}{\bibfnamefont{A.}~\bibnamefont{Ding}},
  \bibinfo{author}{\bibfnamefont{A.}~\bibnamefont{Zukauskaite}},
  \bibnamefont{and} \bibinfo{author}{\bibfnamefont{O.}~\bibnamefont{Ambacher}},
  \bibinfo{journal}{in IEEE International Ultrasonics Symposium (IUS), (IEEE,
  Glasgow, 2019)} pp. \bibinfo{pages}{2588--2591} (\bibinfo{year}{2019}).

\bibitem[{\citenamefont{Tasnadi et~al.}(2010)\citenamefont{Tasnadi, Alling,
  H\"oglund, Wingqvist, Birch, Hultman, and Abrikosov}}]{Tasnadi2010}
\bibinfo{author}{\bibfnamefont{F.}~\bibnamefont{Tasnadi}},
  \bibinfo{author}{\bibfnamefont{B.}~\bibnamefont{Alling}},
  \bibinfo{author}{\bibfnamefont{C.}~\bibnamefont{H\"oglund}},
  \bibinfo{author}{\bibfnamefont{G.}~\bibnamefont{Wingqvist}},
  \bibinfo{author}{\bibfnamefont{J.}~\bibnamefont{Birch}},
  \bibinfo{author}{\bibfnamefont{L.}~\bibnamefont{Hultman}}, \bibnamefont{and}
  \bibinfo{author}{\bibfnamefont{I.~A.} \bibnamefont{Abrikosov}},
  \bibinfo{journal}{Phys. Rev. Lett.} \textbf{\bibinfo{volume}{104}},
  \bibinfo{pages}{137601} (\bibinfo{year}{2010}).

\bibitem[{\citenamefont{H\"oglund et~al.}(2010)\citenamefont{H\"oglund, Birch,
  Alling, Bareno, Czigany, Persson, Wingqvist, Zukauskaite, and
  Hultman}}]{Hoglund2010}
\bibinfo{author}{\bibfnamefont{C.}~\bibnamefont{H\"oglund}},
  \bibinfo{author}{\bibfnamefont{J.}~\bibnamefont{Birch}},
  \bibinfo{author}{\bibfnamefont{B.}~\bibnamefont{Alling}},
  \bibinfo{author}{\bibfnamefont{J.}~\bibnamefont{Bareno}},
  \bibinfo{author}{\bibfnamefont{Z.}~\bibnamefont{Czigany}},
  \bibinfo{author}{\bibfnamefont{P.~O.~A.} \bibnamefont{Persson}},
  \bibinfo{author}{\bibfnamefont{G.}~\bibnamefont{Wingqvist}},
  \bibinfo{author}{\bibfnamefont{A.}~\bibnamefont{Zukauskaite}},
  \bibnamefont{and} \bibinfo{author}{\bibfnamefont{L.}~\bibnamefont{Hultman}},
  \bibinfo{journal}{J. Appl. Phys.} \textbf{\bibinfo{volume}{107}},
  \bibinfo{pages}{123515} (\bibinfo{year}{2010}).

\bibitem[{\citenamefont{Zhang et~al.}(2013)\citenamefont{Zhang, Fu, Holec,
  Humphreys, and Moram}}]{Zhang2013}
\bibinfo{author}{\bibfnamefont{S.}~\bibnamefont{Zhang}},
  \bibinfo{author}{\bibfnamefont{W.~Y.} \bibnamefont{Fu}},
  \bibinfo{author}{\bibfnamefont{D.}~\bibnamefont{Holec}},
  \bibinfo{author}{\bibfnamefont{C.~J.} \bibnamefont{Humphreys}},
  \bibnamefont{and} \bibinfo{author}{\bibfnamefont{M.~A.} \bibnamefont{Moram}},
  \bibinfo{journal}{J. Appl. Phys.} \textbf{\bibinfo{volume}{114}},
  \bibinfo{pages}{243516} (\bibinfo{year}{2013}).

\bibitem[{\citenamefont{Caro et~al.}(2015)\citenamefont{Caro, Zhang, Riekkinen,
  Ylilammi, Moram, Lopez-Acevedo, Molarius, and Laurila}}]{Caro2015}
\bibinfo{author}{\bibfnamefont{M.~A.} \bibnamefont{Caro}},
  \bibinfo{author}{\bibfnamefont{S.}~\bibnamefont{Zhang}},
  \bibinfo{author}{\bibfnamefont{T.}~\bibnamefont{Riekkinen}},
  \bibinfo{author}{\bibfnamefont{M.}~\bibnamefont{Ylilammi}},
  \bibinfo{author}{\bibfnamefont{M.~A.} \bibnamefont{Moram}},
  \bibinfo{author}{\bibfnamefont{O.}~\bibnamefont{Lopez-Acevedo}},
  \bibinfo{author}{\bibfnamefont{J.}~\bibnamefont{Molarius}}, \bibnamefont{and}
  \bibinfo{author}{\bibfnamefont{T.}~\bibnamefont{Laurila}},
  \bibinfo{journal}{J. Phys.: Condens. Matter} \textbf{\bibinfo{volume}{27}},
  \bibinfo{pages}{245901} (\bibinfo{year}{2015}).

\bibitem[{\citenamefont{Zunger et~al.}(1990)\citenamefont{Zunger, Wei,
  Ferreira, and Bernard}}]{Zunger1990}
\bibinfo{author}{\bibfnamefont{A.}~\bibnamefont{Zunger}},
  \bibinfo{author}{\bibfnamefont{S.~H.} \bibnamefont{Wei}},
  \bibinfo{author}{\bibfnamefont{L.~G.} \bibnamefont{Ferreira}},
  \bibnamefont{and} \bibinfo{author}{\bibfnamefont{J.~E.}
  \bibnamefont{Bernard}}, \bibinfo{journal}{Phys. Rev. Lett.}
  \textbf{\bibinfo{volume}{65}}, \bibinfo{pages}{353} (\bibinfo{year}{1990}).

\bibitem[{PWS()}]{PWSCF}
\bibinfo{howpublished}{https://www.quantum-espresso.org/}.

\bibitem[{\citenamefont{Giannozzi et~al.}(2009)\citenamefont{Giannozzi, Baroni,
  Bonini, Calandra, Car, Cavazzoni, Ceresoli, Chiarotti, Cococcioni, Dabo
  et~al.}}]{Giannozzi2009}
\bibinfo{author}{\bibfnamefont{P.}~\bibnamefont{Giannozzi}},
  \bibinfo{author}{\bibfnamefont{S.}~\bibnamefont{Baroni}},
  \bibinfo{author}{\bibfnamefont{N.}~\bibnamefont{Bonini}},
  \bibinfo{author}{\bibfnamefont{M.}~\bibnamefont{Calandra}},
  \bibinfo{author}{\bibfnamefont{R.}~\bibnamefont{Car}},
  \bibinfo{author}{\bibfnamefont{C.}~\bibnamefont{Cavazzoni}},
  \bibinfo{author}{\bibfnamefont{D.}~\bibnamefont{Ceresoli}},
  \bibinfo{author}{\bibfnamefont{G.~L.} \bibnamefont{Chiarotti}},
  \bibinfo{author}{\bibfnamefont{M.}~\bibnamefont{Cococcioni}},
  \bibinfo{author}{\bibfnamefont{I.}~\bibnamefont{Dabo}}, \bibnamefont{et~al.},
  \bibinfo{journal}{J. Phys.: Condens. Matter} \textbf{\bibinfo{volume}{21}},
  \bibinfo{pages}{395502} (\bibinfo{year}{2009}).

\bibitem[{\citenamefont{Prandini et~al.}(2018)\citenamefont{Prandini, Marrazzo,
  Castelli, Mounet, and Marzari}}]{Prandini2018}
\bibinfo{author}{\bibfnamefont{G.}~\bibnamefont{Prandini}},
  \bibinfo{author}{\bibfnamefont{A.}~\bibnamefont{Marrazzo}},
  \bibinfo{author}{\bibfnamefont{I.~E.} \bibnamefont{Castelli}},
  \bibinfo{author}{\bibfnamefont{N.}~\bibnamefont{Mounet}}, \bibnamefont{and}
  \bibinfo{author}{\bibfnamefont{N.}~\bibnamefont{Marzari}},
  \emph{\bibinfo{title}{A standard solid state pseudopotentials (sssp) library
  optimized for accuracy and efficiency (version 1.0)}},
  \bibinfo{howpublished}{Materials Cloud Archive, doi:
  10.24435/materialscloud:2018.0001/} (\bibinfo{year}{2018}).

\bibitem[{\citenamefont{Lejaeghere et~al.}(2016)\citenamefont{Lejaeghere,
  Bihlmayer, Bj\"orkman, Blaha, Bl\"ugel, Blum, Caliste, Castelli, Clark,
  Dal~Corso et~al.}}]{Lejaeghere2016}
\bibinfo{author}{\bibfnamefont{K.}~\bibnamefont{Lejaeghere}},
  \bibinfo{author}{\bibfnamefont{G.}~\bibnamefont{Bihlmayer}},
  \bibinfo{author}{\bibfnamefont{T.}~\bibnamefont{Bj\"orkman}},
  \bibinfo{author}{\bibfnamefont{P.}~\bibnamefont{Blaha}},
  \bibinfo{author}{\bibfnamefont{S.}~\bibnamefont{Bl\"ugel}},
  \bibinfo{author}{\bibfnamefont{V.}~\bibnamefont{Blum}},
  \bibinfo{author}{\bibfnamefont{D.}~\bibnamefont{Caliste}},
  \bibinfo{author}{\bibfnamefont{I.~E.} \bibnamefont{Castelli}},
  \bibinfo{author}{\bibfnamefont{S.~J.} \bibnamefont{Clark}},
  \bibinfo{author}{\bibfnamefont{A.}~\bibnamefont{Dal~Corso}},
  \bibnamefont{et~al.}, \bibinfo{journal}{Science}
  \textbf{\bibinfo{volume}{351}}, \bibinfo{pages}{1415} (\bibinfo{year}{2016}).

\bibitem[{\citenamefont{Grau-Crespo et~al.}(2007)\citenamefont{Grau-Crespo,
  Hamad, Catlow, and de~Leeuw}}]{Grau2007}
\bibinfo{author}{\bibfnamefont{R.}~\bibnamefont{Grau-Crespo}},
  \bibinfo{author}{\bibfnamefont{S.}~\bibnamefont{Hamad}},
  \bibinfo{author}{\bibfnamefont{C.~R.~A.} \bibnamefont{Catlow}},
  \bibnamefont{and} \bibinfo{author}{\bibfnamefont{N.~H.}
  \bibnamefont{de~Leeuw}}, \bibinfo{journal}{J. Phys.: Condens. Matter}
  \textbf{\bibinfo{volume}{19}}, \bibinfo{pages}{256201}
  (\bibinfo{year}{2007}).

\bibitem[{\citenamefont{Farrer and Bellaiche}(2002)}]{Farrer2002}
\bibinfo{author}{\bibfnamefont{N.}~\bibnamefont{Farrer}} \bibnamefont{and}
  \bibinfo{author}{\bibfnamefont{L.}~\bibnamefont{Bellaiche}},
  \bibinfo{journal}{Phys. Rev. B} \textbf{\bibinfo{volume}{66}},
  \bibinfo{pages}{201203(R)} (\bibinfo{year}{2002}).

\bibitem[{\citenamefont{Dridi et~al.}(2003)\citenamefont{Dridi, Bouhafs, and
  Ruterana}}]{Dridi2003}
\bibinfo{author}{\bibfnamefont{Z.}~\bibnamefont{Dridi}},
  \bibinfo{author}{\bibfnamefont{B.}~\bibnamefont{Bouhafs}}, \bibnamefont{and}
  \bibinfo{author}{\bibfnamefont{P.}~\bibnamefont{Ruterana}},
  \bibinfo{journal}{Semicond. Sci. Technol.} \textbf{\bibinfo{volume}{18}},
  \bibinfo{pages}{850–} (\bibinfo{year}{2003}).

\bibitem[{\citenamefont{Ambacher et~al.}(2002)\citenamefont{Ambacher, Majewski,
  Miskys, Link, Hermann, Eick-hoff, Stutzmann, Bernardini, Fiorentini, Tilak
  et~al.}}]{Ambacher2002}
\bibinfo{author}{\bibfnamefont{O.}~\bibnamefont{Ambacher}},
  \bibinfo{author}{\bibfnamefont{J.}~\bibnamefont{Majewski}},
  \bibinfo{author}{\bibfnamefont{C.}~\bibnamefont{Miskys}},
  \bibinfo{author}{\bibfnamefont{A.}~\bibnamefont{Link}},
  \bibinfo{author}{\bibfnamefont{M.}~\bibnamefont{Hermann}},
  \bibinfo{author}{\bibfnamefont{M.}~\bibnamefont{Eick-hoff}},
  \bibinfo{author}{\bibfnamefont{M.}~\bibnamefont{Stutzmann}},
  \bibinfo{author}{\bibfnamefont{F.}~\bibnamefont{Bernardini}},
  \bibinfo{author}{\bibfnamefont{V.}~\bibnamefont{Fiorentini}},
  \bibinfo{author}{\bibfnamefont{V.}~\bibnamefont{Tilak}},
  \bibnamefont{et~al.}, \bibinfo{journal}{J. Phys.: Condens. Matter}
  \textbf{\bibinfo{volume}{14}}, \bibinfo{pages}{3399} (\bibinfo{year}{2002}).

\bibitem[{\citenamefont{Momida et~al.}(2016)\citenamefont{Momida, Teshigahara,
  and Oguchi}}]{Momida2016}
\bibinfo{author}{\bibfnamefont{H.}~\bibnamefont{Momida}},
  \bibinfo{author}{\bibfnamefont{A.}~\bibnamefont{Teshigahara}},
  \bibnamefont{and} \bibinfo{author}{\bibfnamefont{T.}~\bibnamefont{Oguchi}},
  \bibinfo{journal}{AIP Advances} \textbf{\bibinfo{volume}{6}},
  \bibinfo{pages}{065006} (\bibinfo{year}{2016}).

\bibitem[{\citenamefont{Golesorkhtabar
  et~al.}(2013)\citenamefont{Golesorkhtabar, Pavone, Spitaler, Puschnig, and
  Draxl}}]{Golesorkhtabar2013}
\bibinfo{author}{\bibfnamefont{R.}~\bibnamefont{Golesorkhtabar}},
  \bibinfo{author}{\bibfnamefont{P.}~\bibnamefont{Pavone}},
  \bibinfo{author}{\bibfnamefont{J.}~\bibnamefont{Spitaler}},
  \bibinfo{author}{\bibfnamefont{P.}~\bibnamefont{Puschnig}}, \bibnamefont{and}
  \bibinfo{author}{\bibfnamefont{C.}~\bibnamefont{Draxl}},
  \bibinfo{journal}{Comp. Phys. Commun.} \textbf{\bibinfo{volume}{184}},
  \bibinfo{pages}{1861} (\bibinfo{year}{2013}).

\bibitem[{\citenamefont{Lepkowski}(2015)}]{Lepkowski2015}
\bibinfo{author}{\bibfnamefont{S.~P.} \bibnamefont{Lepkowski}},
  \bibinfo{journal}{J. Appl. Phys.} \textbf{\bibinfo{volume}{117}},
  \bibinfo{pages}{105703} (\bibinfo{year}{2015}).

\bibitem[{\citenamefont{Vanderbilt}(2000)}]{Vanderbilt2000}
\bibinfo{author}{\bibfnamefont{D.}~\bibnamefont{Vanderbilt}},
  \bibinfo{journal}{J. Phys. Chem. Solids} \textbf{\bibinfo{volume}{61}},
  \bibinfo{pages}{147} (\bibinfo{year}{2000}).

\bibitem[{\citenamefont{Resta and Vanderbilt}(2007)}]{Resta2007}
\bibinfo{author}{\bibfnamefont{R.}~\bibnamefont{Resta}} \bibnamefont{and}
  \bibinfo{author}{\bibfnamefont{D.}~\bibnamefont{Vanderbilt}}, in
  \emph{\bibinfo{booktitle}{Physics of Ferroelectrics: a Modern Perspective}},
  edited by \bibinfo{editor}{\bibfnamefont{C.~H.} \bibnamefont{Ahn}},
  \bibinfo{editor}{\bibfnamefont{K.~M.} \bibnamefont{Rabe}}, \bibnamefont{and}
  \bibinfo{editor}{\bibfnamefont{J.~M.} \bibnamefont{Triscone}}
  (\bibinfo{publisher}{Springer, Berlin}, \bibinfo{year}{2007}).

\bibitem[{\citenamefont{Bernardini et~al.}(1997)\citenamefont{Bernardini,
  Fiorentini, and Vanderbilt}}]{Bernardini1997}
\bibinfo{author}{\bibfnamefont{F.}~\bibnamefont{Bernardini}},
  \bibinfo{author}{\bibfnamefont{V.}~\bibnamefont{Fiorentini}},
  \bibnamefont{and}
  \bibinfo{author}{\bibfnamefont{D.}~\bibnamefont{Vanderbilt}},
  \bibinfo{journal}{Phys. Rev. B} \textbf{\bibinfo{volume}{56}},
  \bibinfo{pages}{R10024} (\bibinfo{year}{1997}).

\bibitem[{\citenamefont{Kazan et~al.}(2007)\citenamefont{Kazan, Moussaed,
  Nader, and Masri}}]{Kazan2007}
\bibinfo{author}{\bibfnamefont{M.}~\bibnamefont{Kazan}},
  \bibinfo{author}{\bibfnamefont{E.}~\bibnamefont{Moussaed}},
  \bibinfo{author}{\bibfnamefont{R.}~\bibnamefont{Nader}}, \bibnamefont{and}
  \bibinfo{author}{\bibfnamefont{P.}~\bibnamefont{Masri}},
  \bibinfo{journal}{Phys. Status Solidi (c)} \textbf{\bibinfo{volume}{4}},
  \bibinfo{pages}{204} (\bibinfo{year}{2007}).

\bibitem[{\citenamefont{Sotnikov et~al.}(2010)\citenamefont{Sotnikov, Schmidt,
  Weihnacht, Smirnova, Chemekova, and Makarov}}]{Sotnikov2010}
\bibinfo{author}{\bibfnamefont{A.}~\bibnamefont{Sotnikov}},
  \bibinfo{author}{\bibfnamefont{H.}~\bibnamefont{Schmidt}},
  \bibinfo{author}{\bibfnamefont{M.}~\bibnamefont{Weihnacht}},
  \bibinfo{author}{\bibfnamefont{E.}~\bibnamefont{Smirnova}},
  \bibinfo{author}{\bibfnamefont{T.}~\bibnamefont{Chemekova}},
  \bibnamefont{and} \bibinfo{author}{\bibfnamefont{Y.}~\bibnamefont{Makarov}},
  \bibinfo{journal}{IEEE Trans. UFFC} \textbf{\bibinfo{volume}{57}},
  \bibinfo{pages}{808} (\bibinfo{year}{2010}).

\bibitem[{\citenamefont{McNeil et~al.}(1993)\citenamefont{McNeil, Grimsditch,
  and French}}]{McNeil1993}
\bibinfo{author}{\bibfnamefont{L.~E.} \bibnamefont{McNeil}},
  \bibinfo{author}{\bibfnamefont{M.}~\bibnamefont{Grimsditch}},
  \bibnamefont{and} \bibinfo{author}{\bibfnamefont{R.~H.}
  \bibnamefont{French}}, \bibinfo{journal}{J. Am. Ceram. Soc.}
  \textbf{\bibinfo{volume}{76}}, \bibinfo{pages}{1132} (\bibinfo{year}{1993}).

\bibitem[{\citenamefont{Deger et~al.}(1998)\citenamefont{Deger, Born, Angerer,
  Ambacher, Stutzmann, Hormsteiner, Riha, and Fischerauer}}]{Deger1998}
\bibinfo{author}{\bibfnamefont{C.}~\bibnamefont{Deger}},
  \bibinfo{author}{\bibfnamefont{E.}~\bibnamefont{Born}},
  \bibinfo{author}{\bibfnamefont{H.}~\bibnamefont{Angerer}},
  \bibinfo{author}{\bibfnamefont{O.}~\bibnamefont{Ambacher}},
  \bibinfo{author}{\bibfnamefont{M.}~\bibnamefont{Stutzmann}},
  \bibinfo{author}{\bibfnamefont{J.}~\bibnamefont{Hormsteiner}},
  \bibinfo{author}{\bibfnamefont{E.}~\bibnamefont{Riha}}, \bibnamefont{and}
  \bibinfo{author}{\bibfnamefont{G.}~\bibnamefont{Fischerauer}},
  \bibinfo{journal}{Appl. Phys. Lett.} \textbf{\bibinfo{volume}{72}},
  \bibinfo{pages}{2400} (\bibinfo{year}{1998}).

\bibitem[{\citenamefont{Tsubouchi et~al.}(1981)\citenamefont{Tsubouchi, Sugai,
  and Mikoshiba}}]{Tsubouchi1981}
\bibinfo{author}{\bibfnamefont{K.}~\bibnamefont{Tsubouchi}},
  \bibinfo{author}{\bibfnamefont{K.}~\bibnamefont{Sugai}}, \bibnamefont{and}
  \bibinfo{author}{\bibfnamefont{N.}~\bibnamefont{Mikoshiba}},
  \bibinfo{journal}{Proc. IEEE Ultrason. Symp.} pp.
  \bibinfo{pages}{375–--380} (\bibinfo{year}{1981}).

\bibitem[{\citenamefont{de~Jong
  et~al.}(2015{\natexlab{a}})\citenamefont{de~Jong, Chen, Angsten, Jain,
  Notestine, Gamst, Sluiter, Krishna~Ande, van~der Zwaag, Plata
  et~al.}}]{deJong2015a}
\bibinfo{author}{\bibfnamefont{M.}~\bibnamefont{de~Jong}},
  \bibinfo{author}{\bibfnamefont{W.}~\bibnamefont{Chen}},
  \bibinfo{author}{\bibfnamefont{T.}~\bibnamefont{Angsten}},
  \bibinfo{author}{\bibfnamefont{A.}~\bibnamefont{Jain}},
  \bibinfo{author}{\bibfnamefont{R.}~\bibnamefont{Notestine}},
  \bibinfo{author}{\bibfnamefont{A.}~\bibnamefont{Gamst}},
  \bibinfo{author}{\bibfnamefont{M.}~\bibnamefont{Sluiter}},
  \bibinfo{author}{\bibfnamefont{C.}~\bibnamefont{Krishna~Ande}},
  \bibinfo{author}{\bibfnamefont{S.}~\bibnamefont{van~der Zwaag}},
  \bibinfo{author}{\bibfnamefont{J.~J.} \bibnamefont{Plata}},
  \bibnamefont{et~al.}, \bibinfo{journal}{Sci. Data}
  \textbf{\bibinfo{volume}{2}}, \bibinfo{pages}{150009}
  (\bibinfo{year}{2015}{\natexlab{a}}).

\bibitem[{\citenamefont{Wright}(1997)}]{Wright1997}
\bibinfo{author}{\bibfnamefont{A.~F.} \bibnamefont{Wright}},
  \bibinfo{journal}{J. Appl. Phys.} \textbf{\bibinfo{volume}{82}},
  \bibinfo{pages}{2833} (\bibinfo{year}{1997}).

\bibitem[{\citenamefont{Kim et~al.}(1996)\citenamefont{Kim, Lambrecht, and
  Segall}}]{Kim1996}
\bibinfo{author}{\bibfnamefont{K.}~\bibnamefont{Kim}},
  \bibinfo{author}{\bibfnamefont{W.~R.~L.} \bibnamefont{Lambrecht}},
  \bibnamefont{and} \bibinfo{author}{\bibfnamefont{B.}~\bibnamefont{Segall}},
  \bibinfo{journal}{Phys. Rev. B} \textbf{\bibinfo{volume}{53}},
  \bibinfo{pages}{16310} (\bibinfo{year}{1996}).

\bibitem[{\citenamefont{G.~Bu et~al.}(2004)\citenamefont{G.~Bu, Ciplys, Shur,
  Schowalter, Schujman, and Gaska}}]{Bu2004}
\bibinfo{author}{\bibfnamefont{G.}~\bibnamefont{G.~Bu}},
  \bibinfo{author}{\bibfnamefont{D.}~\bibnamefont{Ciplys}},
  \bibinfo{author}{\bibfnamefont{M.}~\bibnamefont{Shur}},
  \bibinfo{author}{\bibfnamefont{L.~J.} \bibnamefont{Schowalter}},
  \bibinfo{author}{\bibfnamefont{S.}~\bibnamefont{Schujman}}, \bibnamefont{and}
  \bibinfo{author}{\bibfnamefont{R.}~\bibnamefont{Gaska}},
  \bibinfo{journal}{Appl. Phys. Lett.} \textbf{\bibinfo{volume}{84}},
  \bibinfo{pages}{4611} (\bibinfo{year}{2004}).

\bibitem[{\citenamefont{Tsubouchi and Mikoshiba}(1985)}]{Tsubouchi1985}
\bibinfo{author}{\bibfnamefont{K.}~\bibnamefont{Tsubouchi}} \bibnamefont{and}
  \bibinfo{author}{\bibfnamefont{N.}~\bibnamefont{Mikoshiba}},
  \bibinfo{journal}{IEEE Trans on Sonics and Ultrasonics}
  \textbf{\bibinfo{volume}{SU-32}}, \bibinfo{pages}{634}
  (\bibinfo{year}{1985}).

\bibitem[{\citenamefont{de~Jong
  et~al.}(2015{\natexlab{b}})\citenamefont{de~Jong, Chen, Geerlings, Asta, and
  Persson}}]{deJong2015b}
\bibinfo{author}{\bibfnamefont{M.}~\bibnamefont{de~Jong}},
  \bibinfo{author}{\bibfnamefont{W.}~\bibnamefont{Chen}},
  \bibinfo{author}{\bibfnamefont{H.}~\bibnamefont{Geerlings}},
  \bibinfo{author}{\bibfnamefont{M.}~\bibnamefont{Asta}}, \bibnamefont{and}
  \bibinfo{author}{\bibfnamefont{K.~A.} \bibnamefont{Persson}},
  \bibinfo{journal}{Sci. Data} \textbf{\bibinfo{volume}{2}},
  \bibinfo{pages}{150053} (\bibinfo{year}{2015}{\natexlab{b}}).

\bibitem[{\citenamefont{Caro et~al.}(2012)\citenamefont{Caro, Schulz, and
  O'Reilly}}]{Caro2012}
\bibinfo{author}{\bibfnamefont{M.~A.} \bibnamefont{Caro}},
  \bibinfo{author}{\bibfnamefont{S.}~\bibnamefont{Schulz}}, \bibnamefont{and}
  \bibinfo{author}{\bibfnamefont{E.~P.} \bibnamefont{O'Reilly}},
  \bibinfo{journal}{J. Phys: Condens. Matter} \textbf{\bibinfo{volume}{25}},
  \bibinfo{pages}{025803} (\bibinfo{year}{2012}).

\end{thebibliography}

\end{document}